\documentclass[12pt,a4paper,aps,preprint,preprintnumbers,superscriptaddress,nofootinbib]{revtex4-1}
\usepackage[english]{babel}
\usepackage[utf8]{inputenc}
\usepackage{graphicx}   
\usepackage{slashed}
\usepackage{epstopdf}
\usepackage{verbatim}   
\usepackage{color}      
\usepackage{multirow}

\usepackage[colorlinks=true, pdfstartview=FitV, linkcolor=blue, citecolor=blue, urlcolor=blue]{hyperref}
\usepackage{float}
\restylefloat{table}
\usepackage{bm}
\usepackage[normalem]{ulem}
\usepackage{amsmath,amssymb}

\def\addresses#1#2{\hbox to \hsize{\@tablebox{#1}\hfil\@tablebox{#2}}}
\def\@tablebox#1{\vtop{\hsize=5in \begin{flushleft} #1 \end{flushleft}}}

\def\beq{\begin{equation}}
\def\eeq{\end{equation}}
\def\bit{\begin{itemize}}
\def\eit{\end{itemize}}
\def\be{\begin{eqnarray}}
\def\ee{\end{eqnarray}}
\def\bray{\begin{array}}
\def\eray{\end{array}}

\definecolor{darkgreen}{rgb}{0,0.6,0}

\definecolor{orange}{rgb}{1,0.5,0}
\definecolor{blue}{rgb}{0,0,1}

\begin{document}
\title{Measurement of Higgs boson self-couplings through $2\rightarrow 3$ vector bosons scattering in future muon colliders}

\author{Junmou Chen}
\email{chenjm@jnu.edu.cn}
\affiliation{Jinan University, Guangzhou, Guangdong 510632, China}

\author{Tong Li}
\email{litong@nankai.edu.cn}
\affiliation{
School of Physics, Nankai University, Tianjin 300071, China
}

\author{Chih-Ting Lu}
\email{timluyu@kias.re.kr}
\affiliation{
School of Physics, Korean Institute for Advanced Study, Seoul, 02455, Korea
}

\author{Yongcheng Wu}
\email{ywu@okstate.edu}
\affiliation{
Department of Physics, Oklahoma State University, Stillwater, Oklahoma, 74078, USA
}

\author{Chang-Yuan Yao}
\email{yaocy@nankai.edu.cn}
\affiliation{
School of Physics, Nankai University, Tianjin 300071, China
}

\begin{abstract}
The $2\rightarrow 3$ scattering of longitudinal vector bosons (VBS) has been proven to be a complementary channel to measure the Higgs self-couplings in the Standard Model (SM) and the SM effective field theory (SMEFT) at colliders. We perform the first comprehensive study of all main $2\rightarrow 3$ VBS processes at high-energy muon colliders, especially including background analysis. The main contributing channels turn out to be the scattering of $W^+W^-\rightarrow W^+W^-h$, $ZZh$ and $hhh$. We obtain the constraints on $c_6$ and $c_{\Phi_1}$ which are the Wilson coefficients of the dimension-6 operators relevant for Higgs self-couplings in the SMEFT.
With the center-of-mass energies of 10 TeV and 30 TeV, we find the expected sensitivity to the coefficients $c_6/\Lambda^2$ and $c_{\Phi_1}/\Lambda^2$ can reach the level of 0.01 TeV$^{-2}$. The tightest constraints come from the $hhh$ channel, while the constraints from $WWh$ are also comparable. Our results crucially depend on selecting the longitudinal polarizations for the final $W$ and $Z$ bosons. We then study how the sensitivities change by varying the efficiency of tagging longitudinal polarizations, and find that the significance remains consistently high.
\end{abstract}


\maketitle

\section{Introduction}

The self-couplings of the Standard Model (SM) Higgs boson are most closely related to the origin of the electroweak (EW) symmetry breaking. Any deviation away from the SM prediction indicates the existence of new physics above the EW scale~\cite{Higgs:1964pj,Englert:1964et,Guralnik:1964eu,Agrawal:2019bpm,DiMicco:2019ngk} and may shed light on the electroweak baryogenesis~\cite{Kuzmin:1985mm,Cohen:1993nk,Rubakov:1996vz,Morrissey:2012db,Huang:2015izx,Huang:2015tdv,Reichert:2017puo,Kanemura:2020yyr}. The precision measurement of the Higgs self-couplings is thus one of the most important topics of high-energy physics, as well as one of the main goals of the Large Hadron Collider (LHC) upgrades~\cite{Cepeda:2019klc} and future high-energy lepton colliders~\cite{Delahaye:2019omf,Long:2020wfp,Buttazzo:2018qqp,Costantini:2020stv,Han:2020pif}.

The traditional channel of measuring Higgs self-couplings is well established through the Higgs pair production via gluon fusion at hadron colliders~\cite{Baur:2003gpa,Baur:2003gp,Goertz:2014qta,Azatov:2015oxa}.
In recent years, people have been proposing and
studying the measurement of Higgs couplings through the processes involving longitudinal
vector bosons~\cite{Giudice:2007fh,Dolan:2013rja,Ling:2014sne,Cao:2015oxx,Bishara:2016kjn,Stolarski:2020qim}.
The most common channels of measuring Higgs self-couplings through longitudinal vector bosons are the $2\rightarrow 2$ vector boson scattering (VBS) because they have the largest production cross sections. However, it was first proposed in Ref.~\cite{Henning:2018kys} that the $2\rightarrow 3$ VBS can also be used to probe Higgs self-couplings
\begin{eqnarray}
V_LV_L\to V_L V_L h\;,
\end{eqnarray}
where $V_L$ denotes the longitudinally polarized gauge bosons and $h$ is the physical SM Higgs boson.
In fact, the VBS $V_LV_L\to V_L V_L h$ belongs to the more general proposal of studying the measurement of Higgs couplings through the processes involving longitudinal vector bosons~\cite{Henning:2018kys, Costantini:2020stv,Chen:2021rid}. This approach is based on the Goldstone equivalence theorem (GET), stating that the scattering amplitudes of the longitudinal vector bosons are equivalent to the amplitudes of the corresponding Goldstone bosons. After taking the GET, the structure of $SU(2)\times U(1)$ group for the Higgs doublet ensures that the couplings of Goldstone bosons and the Higgs couplings come from the same parameters. This is in particular true for the SM as well as the SM effective field theory (SMEFT)~\cite{Grzadkowski:2010es} which is a valuable tool for parametrizing new physics in much higher energies.

For the $2\rightarrow 3$ VBS, the dimension-6 operator $(\Phi^\dag\Phi)^3$ which is directly related to the shape of Higgs potential can provide optimistic probing channels for the Higgs self-coupling measurement. After taking extra care of choosing longitudinal polarizations and applying advanced transverse momentum cuts, a very naive estimation shows that the constraints on the Higgs self-couplings could be comparable to di-Higgs production in future high-energy colliders~\cite{Henning:2018kys,Chen:2021rid}. This might be surprising at first sight, as the cross section of $2\rightarrow 3$ VBS is much smaller than that of $2\rightarrow 2$ VBS. However, according to the GET, the amplitudes beyond the SM $\mathcal{A}^{\text{BSM}}$ in $2\rightarrow 3$ VBS demonstrate quadratic energy increase compared to the amplitude in the SM $\mathcal{A}^{\text{SM}}$, i.e., $\mathcal{A}^{\text{BSM}}/\mathcal{A}^{\text{SM}} \simeq E^2/\Lambda^2$
where $E$ denotes the collision energy and $\Lambda$ is the new physics scale.
As a result, the sensitivity to new physics from the BSM amplitudes balances out the small cross sections in $2\rightarrow 3$ VBS and  the future high-energy colliders demonstrate extremely high sensitivity to new physics.
In Ref.~\cite{Chen:2021rid}, some of us studied the $2\rightarrow 3$ processes with either $V_LV_Lh$ or $hhh$ final states at the LHC and future lepton colliders and demonstrated extremely high sensitivity to new physics.
However, in Ref.~\cite{Chen:2021rid} we did not consider the decay of the final states, nor did we carry out the detailed background analysis. Moreover, we only analyzed the productions of $W^+W^-h~(W^{\pm}W^{\pm}h)$ and $hhh$ without considering other similar channels.

The $V_LV_L$ initiated VBS prefers dynamically activated EW gauge bosons at high-energy colliders. Recently, with new technological development, the high-energy muon colliders have again gained much attention in the community due to the extremely low synchrotron radiation and much smaller beam-energy spread~\cite{MICE:2019jkl,Delahaye:2019omf,Bartosik:2020xwr}.
A muon collider operating at much higher energies above multi-TeV offers great potential in reaching new physics up to its energy threshold.
Another advantage of the muon collider is that it can reach a high integrated luminosity scaling with energy quadratically and does not suffer from QCD background. Thus, the muon collider provides good opportunities to precisely measure the Higgs self-couplings through the VBS~\cite{Han:2020pif,Costantini:2020stv,AlAli:2021let}.
In this paper, we carry out a comprehensive study of all $2\rightarrow 3$ VBS processes that are related to Higgs self-couplings as well as the detailed background analysis at high-energy muon colliders. We aim to obtain a reliable estimation of the sensitivity to the related Wilson coefficients of dim-6 operators in SMEFT.

The rest of the paper is organized as follows. In Sec.~\ref{sec:SMEFT}, we briefly describe the framework of SMEFT and emphasize the operators relevant for the Higgs self-couplings. We also summarize the characteristics and advantages of the $2\rightarrow 3$ VBS. We discuss the implementation of the $2\rightarrow 3$ VBS analysis in Sec.~\ref{sec:sim}. Three VBS topologies are analyzed at muon collider, including $WW\to WWh, ZZh, hhh$, and we show the projected sensitivity to the Higgs self-couplings.
Finally, we summarize our conclusions in Sec.~\ref{sec:Con}.

\section{The $2\rightarrow 3$ VBS under the framework of SMEFT}
\label{sec:SMEFT}

In this section we give an overview about the underlying physics behind the measurement of Higgs self-couplings through $2\rightarrow 3$ VBS~\cite{Chen:2021rid}. Based on this overview, we then enumerate all possible $2\rightarrow 3$ VBS that can be used for Higgs self-coupling measurement, and compare with other processes such as $2\rightarrow 2$ VBS, $2\rightarrow 4$ VBS and etc.

According to the GET, the amplitudes of the longitudinal vector boson scatterings are approximately equal to those of the corresponding Goldstone bosons' scattering. Moreover, the Goldstone bosons and the Higgs boson form an $SU(2)$ Higgs doublet in the SM and SMEFT. They thus couple to other SM particles together with the same couplings. This characteristic means that any Higgs couplings can be measured through the production of either the Higgs boson or the Goldstone bosons, as long as the $SU(2)$ gauge symmetry is preserved or spontaneously broken. Note that, with large energy scale $E\gg M_Z$, the SM gauge symmetry is efficiently restored in high-energy collisions~\cite{Chen:2016wkt,Bauer:2017isx,Bauer:2018arx,Han:2020uid}. The EW gauge bosons in the unbroken SM are
dynamically activated. The high-energy leptonic colliders thus provide an ideal environment to make use of the GET and measure the Higgs self-couplings through the scattering of the longitudinally polarized gauge bosons.

For the $2\rightarrow 3$ VBS under SMEFT, the relevant dim-6 operators involved in the measurement of Higgs self-couplings include
\begin{eqnarray}
\label{eq:EFT_operator}
{\mathcal L}_{\text{dim}-6}\supset \frac{1}{\Lambda^2}\left[c_6\mathcal{O}_6+c_{\Phi_1}\mathcal{O}_{\Phi_1} \right]= \frac{1}{\Lambda^2}\left[ c_6 (\Phi^\dag\Phi)^3 + c_{\Phi_1} \partial^{\mu}(\Phi^\dag\Phi)\partial_{\mu}(\Phi^\dag\Phi)\right]\;,
\end{eqnarray}
where $\Phi$ is the SM Higgs doublet and can be parametrized as
\begin{align}
\Phi=\left(\begin{array}{c}
    \phi^+ \\
    \frac{v+h+i\phi^0}{\sqrt{2}}
    \end{array}\right)\;.
\end{align}
Here $\phi^+$ and $\phi^0$ denote the charged and neutral Goldstone bosons, respectively.
Given the GET, the $2\rightarrow 3$ scattering $V_LV_L\rightarrow V_LV_Lh$ can be approximated by $\phi\phi\rightarrow \phi\phi h$ with $\phi$ being $\phi^\pm$ or $\phi^0$.
The amplitudes of the above $2\rightarrow 3$ VBS processes at high energies can be easily estimated by taking the GET (see the details in Ref.~\cite{Chen:2021rid}). First of all, the SM amplitudes are suppressed by the propagators at high energies, resulting in $\mathcal{A}^{\text{SM}}\simeq\frac{v}{E^2}$. Since $\mathcal{O}_{\Phi_1}$ gives derivative couplings scaling as $\propto E^2$, the BSM amplitudes give $\mathcal{A}_{\Phi_1}^{\text{BSM}}\simeq \frac{c_{\Phi_1}v}{\Lambda^2}$ after combining with the suppressing effect of the propagators.
The derivation for $c_6$ is different, but leads to the same conclusion. For the $2\rightarrow 3$ VBS from $\mathcal{O}_6$, there is a contact diagram from 5-point scalar vertices contributing to the BSM amplitude that is for instance $\phi^+\phi^-\rightarrow \phi^+\phi^- h$. This contribution stays constant  even as the energy grows. Thus the BSM amplitude for $c_6$ simply becomes $\mathcal{A}_{6}^{\text{BSM}}\simeq \frac{c_6v}{\Lambda^2}$. As a result, we have the approximate ratio of the SM and BSM amplitudes
\begin{eqnarray}
\frac{\mathcal{A}^{\text{BSM}}}{\mathcal{A}^{\text{SM}}} \simeq \frac{E^2}{\Lambda^2}\;,
\label{eq:amp_2to3}
\end{eqnarray}
for both $c_6$ and $c_{\Phi_1}$ coefficients. This behavior indicates the dominant sensitivity of BSM amplitudes over SM ones at high-energy colliders.

The high sensitivity of the BSM amplitudes of $2\rightarrow 3$ VBS over the Wilson coefficients (in particular the $c_6$ coefficient) is the key to understanding the potential importance in measuring Higgs self-couplings in future high-energy colliders.
Although the $2\rightarrow 3$ VBS processes have relatively small production cross sections, the energy behavior Eq.~(\ref{eq:amp_2to3}) becomes an important advantage compared with $2\rightarrow 2$ VBS. In the $2\rightarrow 2$ VBS, the contact diagram has the contributions from both $\mathcal{O}_{\Phi_1}$ and $\mathcal{O}_6$, as well as SM. As a result, the dependence on $c_6$ becomes $\frac{\mathcal{A}^{\text{BSM}}}{\mathcal{A}^{\text{SM}}}\simeq v^2/\Lambda^2$, without any energy enhancement. Also, the dependence on $c_{\Phi_1}$ scales as $E^2$ which is much larger than that of $c_6$. This gives more obstacles to the measurement of $c_6$ which is more directly related to the shape of the Higgs potential and the EW symmetry breaking mechanism. By contrast, the contact diagrams in $2\rightarrow 3$ VBS only arise from $\mathcal{O}_6$ operator, while the dependence on $c_{\Phi_1}$ is suppressed by the propagators. As a result, the amplitude's dependence on $c_6$ is of the same order as that of $c_{\Phi_1}$. This makes it easier to obtain the constraints on $c_6$ in $2\rightarrow 3$ VBS than $2\rightarrow 2$ VBS.
On the other hand, the $2\rightarrow 4$ VBS has very similar energy behavior as Eq.~(\ref{eq:amp_2to3}). However, it has even much smaller cross sections which makes the measurement impractical even in future high-energy colliders with $E\sim \mathcal{O}(10)$ TeV.

From the above analysis, we can also enumerate all $2\rightarrow 3$ VBS processes which are relevant to the measurement of $c_6$, i.e. the Higgs potential. The contact diagram is generated by 5-point scalar vertices which can only be obtained when the vertex has either one Higgs boson, three Higgs bosons, or five Higgs bosons. The $hh\rightarrow hhh$ process cannot be implemented in a leptonic collider. Consequently, the $2\rightarrow 3$ VBS processes we are interested in can be summarized in the following list according to the final states
\begin{eqnarray}
&&W^+W^-/ZZ\rightarrow W^+W^-h\;,~~W^{\pm}W^{\pm}\rightarrow W^{\pm}W^{\pm}h\;,\nonumber \\
&&W^+W^-/ZZ\rightarrow ZZh\;,~~
W^{\pm}Z\rightarrow W^{\pm}Zh\;,\nonumber\\
&&W^+W^-/ZZ\rightarrow hhh\;.
\end{eqnarray}
As we can see, in principle, all processes except for the same-sign $W$ processes can be implemented in $\mu^+\mu^-$ colliders, while the same-sign $W$ processes need the performance of $\mu^{\pm}\mu^{\pm}$ collider.

\section{The numerical simulation and analysis of the $2\rightarrow 3$ VBS at muon colliders}
\label{sec:sim}

The VBF topology for $2\to 3$ productions reads as follows
\begin{eqnarray}
\mu^+ \mu^-\to \nu_\mu \bar{\nu}_\mu V_LV_Lh/hhh~~~(WW~{\rm fusion})\;,\nonumber\\
\mu^+ \mu^-\to \mu^+ \mu^- V_LV_Lh/hhh~~~(ZZ~{\rm fusion})\;.\nonumber
\end{eqnarray}
At high energies the outgoing muons in $ZZ$ (or $WZ$) fusion are extremely forward with a small polar angle and most likely escape the detector. We require the capability of detecting very energetic muons in
the forward region of a few degrees with respect to the beam~\cite{Han:2020pif}.
Thus, the $WW$ and $ZZ$ fusion processes are distinguishable and the $ZZ$ fusions are substantially suppressed by requiring to detect the forward muons~\cite{Han:2020pif}. As a result, we only consider the exclusive $WW$ fusion channels below and give a conservative estimate of the expected sensitivity for the couplings.
The VBF productions are simulated using MadGraph5\_aMC$@$NLO~\cite{Alwall:2014hca}. We make use of the SMEFT UFO model~\cite{Degrande:2020evl} to implement the dim-6 SMEFT operators and allow the decays of the gauge bosons and the Higgs boson.
We set the decay widths of massive particles (e.g. $W^{\pm}$, $Z$ and $h$) to zero and impose a mild cut $m_{\nu\bar{\nu}}>150$ GeV in order to achieve nonphysical cancellation resulting from longitudinal polarization vectors~\cite{Chen:2021rid}. The polarizations of vector bosons in the final states are specified to be longitudinal~\cite{BuarqueFranzosi:2019boy}. The decays of gauge bosons and Higgs boson are implemented in MadSpin~\cite{Artoisenet:2012st}~\footnote{We launch the processes in MadGraph and set the decay widths to zero in the parameter card, instead of importing no$\_$width mode directly. In this way, the main processes are implemented with zero decay widths, while MadSpin can still decay the particles with the original decay widths.}.

Before carrying out background analysis, we apply two kinematic means to make sure that the sensitivity to BSM physics manifested in Eq.~(\ref{eq:amp_2to3}) at the amplitude level also works at the level of cross sections. The first one is to apply significant $p_T$ cuts on the decay products of $W, Z, h$ bosons to reduce the SM cross sections enhanced by infrared singularities.
The second approach, as mentioned above, is to require the polarizations of the $W$ and $Z$ bosons to be longitudinal. As a result, the sensitivities of cross sections to BSM physics are not overwhelmed by the contributions of transverse polarizations (e.g. $W_TW_T\rightarrow W_TW_T$) which are dominant over pure longitudinal polarizations.
In experiments the measurement of polarizations is achievable through reconstructing distinct kinematic properties of the decaying products~\cite{Ballestrero:2017bxn, Ballestrero:2020qgv, De:2020iwq,ATLAS:2019bsc,CMS:2020ezf}. See also Refs.~\cite{Grossi:2020orx, Kim:2021gtv} for recent advance.
Fortunately, the measurements at LHC in recent years shows that the efficiency for tagging longitudinal/transverse polarizations can be quite high, e.g., reaching $70\%-80\%$ in Ref.~\cite{ATLAS:2019bsc} and as high as $99\%$ in Ref.~\cite{CMS:2020ezf}.
When doing the analysis for $2\rightarrow 3$ VBS, we will scan the tagging efficiency of polarizations for $W$ and $Z$ at a reasonable range and compare the corresponding significance and constraints accordingly.
The ratios of longitudinal and transverse polarizations in a process are determined by fitting the parameters in the angular distribution of cross section, supplemented by unitarity. The (mis-)tagging efficiency is obtained in terms of the determined fractions as well as the errors. According to Refs.~\cite{ATLAS:2019bsc,CMS:2020ezf}, the efficiency for the transverse polarization is as high as that for the longitudinal polarization.
The corresponding mistagging efficiencies for the transverse polarization gauge bosons would be smaller than 10\%-20\%. We neglect the contribution originating from transversely polarized gauge bosons wrongly identified as longitudinal polarization in the following analysis.

In this work, we choose two typical energies and integrated luminosities of the muon collider as our benchmarks
\begin{eqnarray}
    \sqrt{s}=10~\text{and}~30~\text{TeV},\quad \mathcal{L}=10~\text{and}~90~ \rm{ab}^{-1}\,,
\end{eqnarray}
corresponding to an optimistic energy quadratic scaling~\cite{Delahaye:2019omf}
\begin{eqnarray}
\mathcal{L}=\Big({\sqrt{s}\over 10~{\rm TeV}}\Big)^2~10~{\rm ab}^{-1}\;.
\end{eqnarray}
The significance is defined as follows
\begin{equation}\label{eq:signif_real}
\mathcal{S}=\sqrt{2 \left[(B+S) \ln (1+S/B)-S\right]}\,,
\end{equation}
where the signal ($S$) and background ($B$) events are given by
\begin{equation}
    S=N_{\text{sig}}-N_{c=0}\,,\quad B=N_{c=0}+N_{\text{bkg}}\,.
\label{eq:SandB}
\end{equation}
The signal events $N_{\rm sig}$ is obtained by setting certain high-dimension coefficient(s) in SMEFT. The background events are composed of $N_{c=0}$ and $N_{\rm bkg}$ in which $N_{c=0}$ denotes the SM prediction with all high-dimension coefficients vanishing and $N_{\rm bkg}$ corresponds to other SM processes giving the same final states. Our Eq.~(\ref{eq:amp_2to3}) signifies the dominance of BSM amplitudes over SM ones for $2\rightarrow 3$ VBS processes at high energies. Thus we have a good reason to expect similar sensitivity in cross sections and local significance. In the case that the systematic error is dominant, the significance is given by $S/B\simeq |\mathcal{A^{BSM}}|^2/|\mathcal{A^{SM}}|^2\simeq  E^4/\Lambda^4$. Then we do have the conclusion of high significance at high energies.
In the case that the statistical error dominates, the local significance can be determined as $S/\sqrt{B}\propto |\mathcal{A^{BSM}}|^2/|\mathcal{A^{SM}}|\simeq E^4/\Lambda^2$.
The energy dependence in this result is similar to the case of $S/B$. We take into account a small systematic uncertainty as 0.1\% according to Ref.~\cite{Han:2020uak}. In Eq.~(\ref{eq:SandB}), we take the signal events as the difference from the SM due to the presence of EFT couplings and also take into account additional SM processes giving the same final states in the background events. The real significance in our analysis is thus more complicated than the above estimation. Nevertheless, our results below demonstrate that our initial estimation in Eq.~(\ref{eq:amp_2to3}) is still correct at least qualitatively.

\subsection{$WW\to WWh$}
\label{wwh}

\begin{figure}[th!]
\centering
\includegraphics[width=0.45\textwidth]{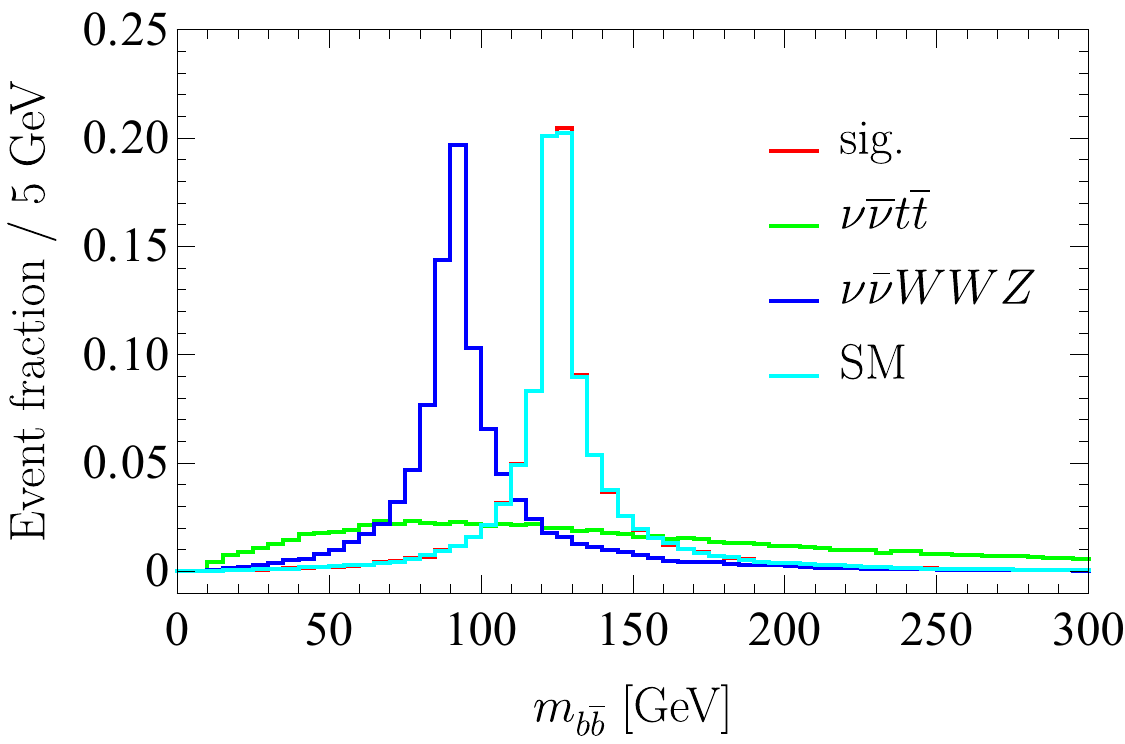}
\includegraphics[width=0.45\textwidth]{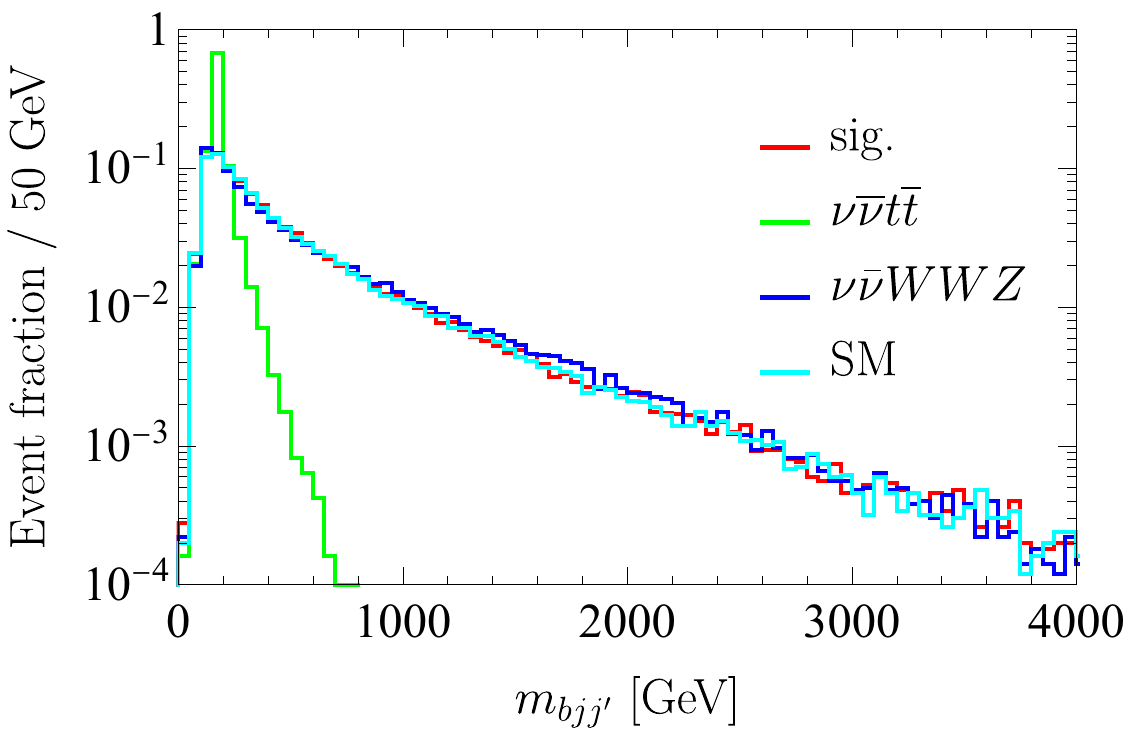}
\includegraphics[width=0.45\textwidth]{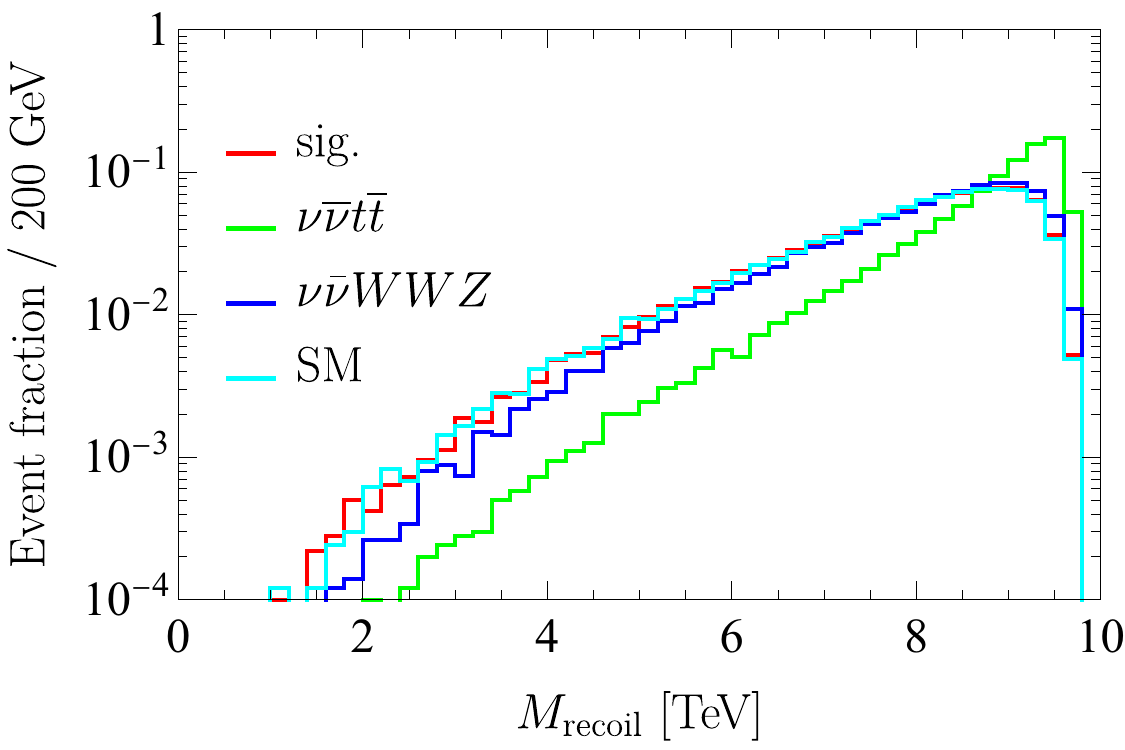}
\caption{The kinematic distributions of $WWh$ signal and SM backgrounds for $\sqrt{s}=10$ TeV. Top left: the invariant mass of the two $b$ jets $m_{b\bar{b}}$. Top right: the reconstructed top mass of $m_{bjj'}$. Bottom: the recoil mass $M_{\rm recoil}$.}
\label{fig:wwh}
\end{figure}

The VBF topology for $WWh$ production reads as follows
\begin{eqnarray}
\mu^+ \mu^-\to \nu_\mu \bar{\nu}_\mu W^+ W^- h~~~(WW~{\rm fusion})\;.
\end{eqnarray}
We require that one of the $W$ bosons hadronically decays into two jets and the other one decays to $\ell^\pm \nu_\ell$. The Higgs boson $h$ is followed by the leading decay channel $h\to b\bar{b}$. The signal would thus be $\ell^\pm j j' b\bar{b}$ plus large missing energy, with the $b\bar{b}$ ($jj'$) pair being near the Higgs ($W$) boson resonance. The irreducible backgrounds consist of
\begin{eqnarray}
\mu^+\mu^-\to t\bar{t},~~\nu_\mu \bar{\nu}_\mu t\bar{t},~~W^+W^-Z,~~\nu_\mu \bar{\nu}_\mu W^+W^-Z~~{\rm and}~~\gamma\gamma\to t\bar{t}
\end{eqnarray}
followed by the semileptonic decays of the two top quarks or the two $W$ bosons as well as $Z\to b\bar{b}$. For $\sqrt{s}=10$~TeV, we impose the basic acceptance cuts on the objects
\begin{eqnarray}\label{cut:wwh_pt}
p_T(\ell,b,j)>50~{\rm GeV}\;,~~~10^{\circ}<\theta_{\ell,b,j}<170^{\circ}\;.
\end{eqnarray}
We also assume an energy resolution for the jets and leptons~\footnote{We will apply the same energy resolution of Eq.~(\ref{energyresol}) for the analysis of $WW\rightarrow ZZh$ and $WW\rightarrow hhh$
in Sec.~\ref{sec:zzh} and Sec.~\ref{sec:hhh}, respectively.}
\begin{eqnarray}
\Delta E/E=5\%\;.
\label{energyresol}
\end{eqnarray}
In the signal events, we reconstruct the Higgs boson $h$ with the invariant mass of the two $b$ jets as shown in Fig.~\ref{fig:wwh} and apply the invariant mass cut
\begin{eqnarray}\label{cut:wwh_mbb}
m_{b\bar{b}}=m_h\pm 15~{\rm GeV}\;.
\end{eqnarray}
In the analysis here and below, we assume an optimistic $b$-tagging efficiency of 100\%.
To distinguish the backgrounds and the signal, we also define the recoil mass as follows
\begin{equation}
M_{\text{recoil}}=\sqrt{\left(p_{\mu^+}+p_{\mu^-}-\sum_{i}p_{i}^{\text{visible}}\right)^2}=\sqrt{\left(\sum_{k}p_{\nu_k}\right)^2}.
\end{equation}
The final states of $t\bar{t}$, $W^+W^-Z$ and $\gamma\gamma\to t\bar{t}$ backgrounds have only one neutrino from $W$ decay and thus the recoil mass turns out to be near zero, which is very different from the signal. In addition, the $\nu_\mu \bar{\nu}_\mu t\bar{t}$, $\nu_\mu \bar{\nu}_\mu W^+W^-Z$ backgrounds exhibit much higher recoil mass as shown in Fig.~\ref{fig:wwh}. After we apply the recoil mass cut
\begin{equation}\label{cut:wwh_top_mass}
    M_{\text{recoil}}>1000~\rm{GeV}\;,
\end{equation}
only $\nu_\mu \bar{\nu}_\mu t\bar{t}$ and $\nu_\mu \bar{\nu}_\mu W^+W^-Z$ backgrounds are left. Since this cut can effectively remove the events from $t\bar{t}$ and $W^+W^-Z$, we will not consider these two backgrounds in our analysis.
There exists a hadronically decayed top quark in the $\nu_\mu \bar{\nu}_\mu t\bar{t}$ background but not in the signal. We first reconstruct the $W$ from the two jets $jj'$ and then select the $b$ jet which minimizes the separation $\Delta R(W,b)$ to reconstruct the top quark. The invariant mass distribution of the top quark is shown in Fig.~\ref{fig:wwh}, and we apply the cut
\begin{equation}
    |m_{bjj'}-m_t|>100~\rm{GeV}
\end{equation}
to reduce the $\nu_\mu \bar{\nu}_\mu t\bar{t}$ background. The $\nu_\mu \bar{\nu}_\mu W^+W^-Z$ background is highly suppressed by the $b\bar{b}$ invariant mass cut.

\begin{figure}[th!]
\centering
\includegraphics[width=0.45\textwidth]{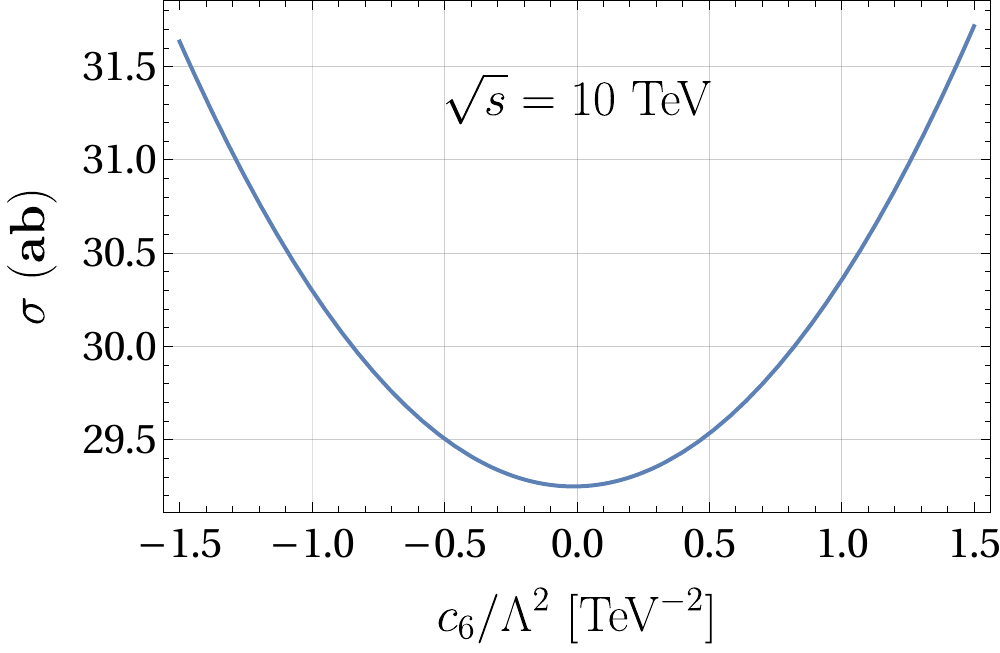}
\includegraphics[width=0.45\textwidth]{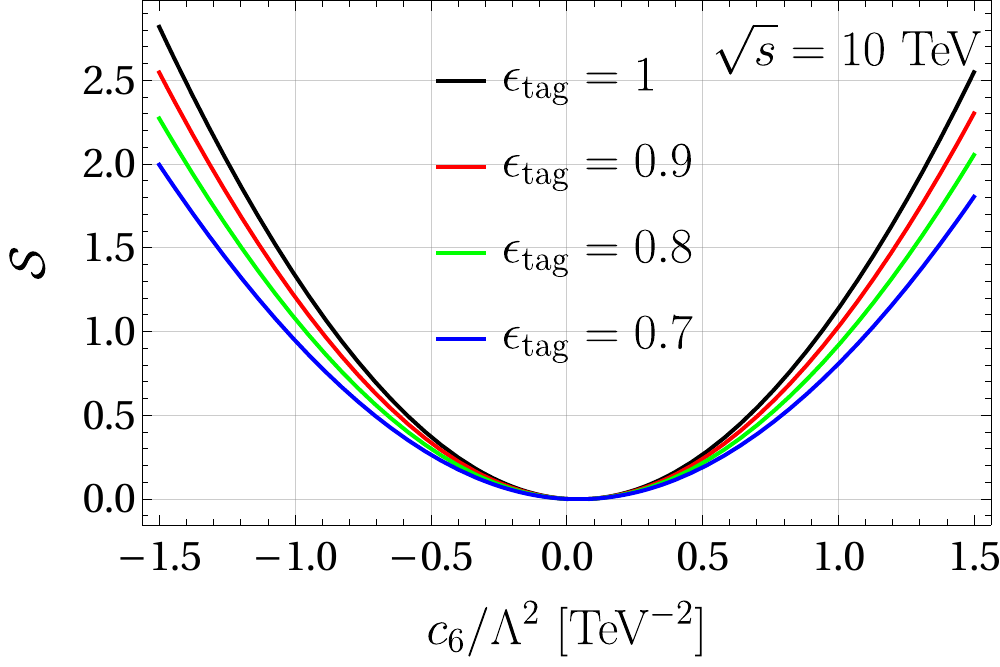}\\
\includegraphics[width=0.45\textwidth]{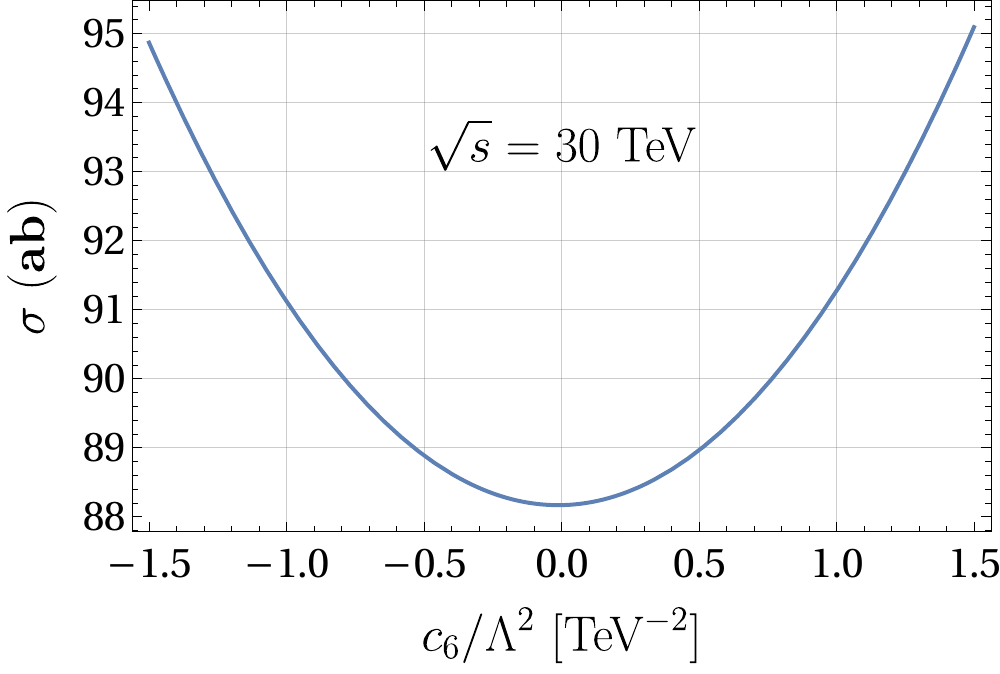}
\includegraphics[width=0.45\textwidth]{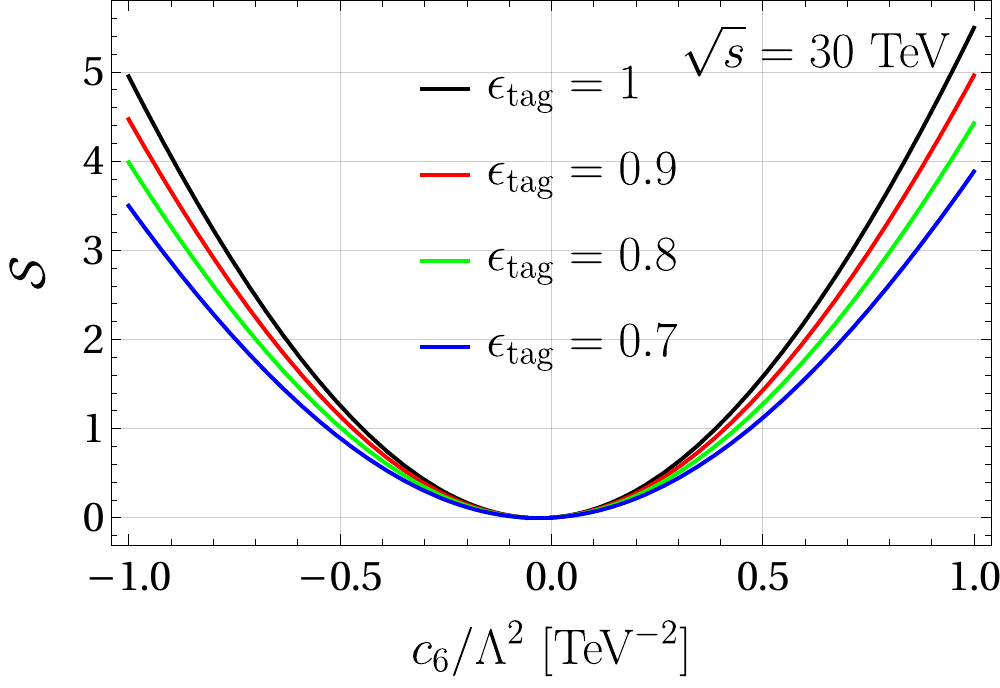}
\caption{The initial cross section of the $WWh$ signal without any cuts (left) and the significance $\mathcal{S}$ after combining the analyzed results (right), as a function of $c_6/\Lambda^2$ for $\sqrt{s}=10$~TeV and $\mathcal{L}=10~\mathrm{ab}^{-1}$ (upper) as well as $\sqrt{s}=30$~TeV and $\mathcal{L}=90~\mathrm{ab}^{-1}$ (lower).}
\label{fig:wwh_significance}
\end{figure}

Next we show the cut efficiencies as well as the number of events after cuts for our $WWh$ signal and backgrounds. For illustration, in Table~\ref{tab:wwh}, we take $c_6/\Lambda^2=-1$ TeV$^{-2}$ and $c_{\Phi_1}/\Lambda^2=0$ and the number of events are given for $\sqrt{s}=10$ TeV and $\mathcal{L}=10$ ab$^{-1}$ outside the brackets. One can obtain the significance as $\mathcal{S}=1.33$.
We then fix $c_{\Phi_1}/\Lambda^2=0$ and perform a scan of the parameter $c_6/\Lambda^2$ in the range $[-1.5,1.5]$ TeV$^{-2}$ to test the projected sensitivity of the effective coupling.
We show the initial cross section of the signal without any cuts (left) and the significance $\mathcal{S}$ after combining the analyzed results (right), as functions of $c_6/\Lambda^2$ in the upper panels of Fig.~\ref{fig:wwh_significance}.
Note that we display the resulting significance under the assumption of various efficiencies $\epsilon_{\rm tag}$ of tagging longitudinal gauge bosons in final states. As expected, higher tagging efficiencies lead to increasing sensitivity.
For $\epsilon_{\rm tag}=1$, the $1\sigma$ and $2\sigma$ sensitivity ranges of $c_6/\Lambda^2$ are $[-0.856, 0.940]$ TeV$^{-2}$ and $[-1.245, 1.327]$ TeV$^{-2}$, respectively. For $\sqrt{s}=30$ TeV and $\mathcal{L}=90$ ab$^{-1}$, we instead require $p_T(\ell,b,j)> 100$~GeV and $M_{\text{recoil}}>4000$~GeV. We also take $c_6/\Lambda^2=-0.5$ TeV$^{-2}$ and $c_{\Phi_1}/\Lambda^2=0$, and show the cut efficiency results inside the brackets in Table~\ref{tab:wwh}. The significance is given as $\mathcal{S}=1.27$ for this setup. One can achieve a better sensitivity of the effective coupling as seen from the lower panels in Fig.~\ref{fig:wwh_significance}. The $1\sigma$ and $2\sigma$ ranges of $c_6/\Lambda^2$ are $[-0.447, 0.389]$ TeV$^{-2}$ and $[-0.627, 0.569]$ TeV$^{-2}$, respectively.

\begin{table}[tb!]
\centering
\begin{tabular}{|c|c|c|c|c|}
\hline
$WWh$ & \text{signal} & $\nu\bar{\nu}t\bar{t}$ & $\nu\bar{\nu}WWZ$ & $c_6=0$ (SM) \\ \hline
initial & 302.703 [7999.836] & 10816.667 [163490.400] & 132.983 [3678.714] & 292.778 [7935.077]  \\ \hline
$p_{T}$ and $\theta$ cut & 21.728 [78.558] & 1701.462 [5071.472] & 6.250 [20.674] & 16.940 [65.067] \\ \hline
$m_{b\bar{b}}$ cut & 14.245 [35.199] & 162.250 [52.316] & 0.894 [2.575] & 11.120 [28.725] \\ \hline
$m_{bjj'}$ cut & 8.960 [26.399] & 0.649 [0.000] & 0.633 [1.912] & 5.270 [20.155] \\ \hline
Efficiency $\epsilon(\%)$ & 2.960 [0.330] & 0.006 [0.000] & 0.476 [0.052] & 1.800 [0.254] \\ \hline
\end{tabular}
\caption{The selection efficiencies for the $WWh$ signal and SM backgrounds. The number of events before and after selection cuts are evaluated with $c_6/\Lambda^2=-1$ TeV$^{-2}$ and $c_{\Phi_1}/\Lambda^2=0$ at muon collider with $\sqrt{s}=10$ TeV and the luminosity of $\mathcal{L}=10$ ab$^{-1}$ (outside brackets), and with $c_6/\Lambda^2=-0.5$ TeV$^{-2}$ and $c_{\Phi_1}/\Lambda^2=0$ at muon collider with $\sqrt{s}=30$ TeV and the luminosity of $\mathcal{L}=90$ ab$^{-1}$ (inside brackets).}
\label{tab:wwh}
\end{table}

\begin{figure}[th!]
\centering
\includegraphics[width=0.45\textwidth]{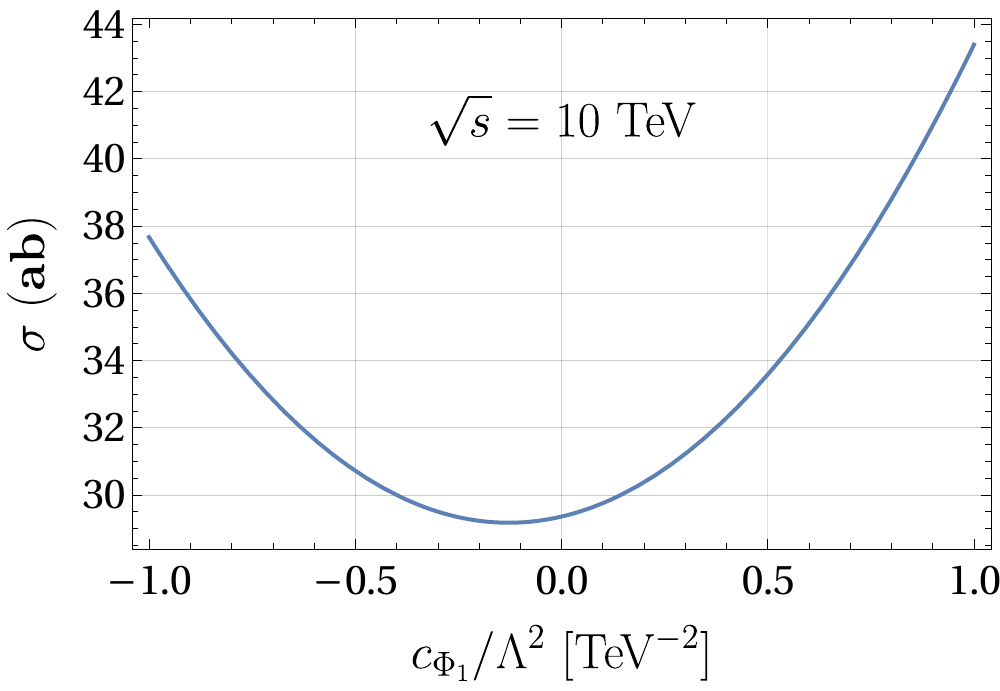}
\includegraphics[width=0.45\textwidth]{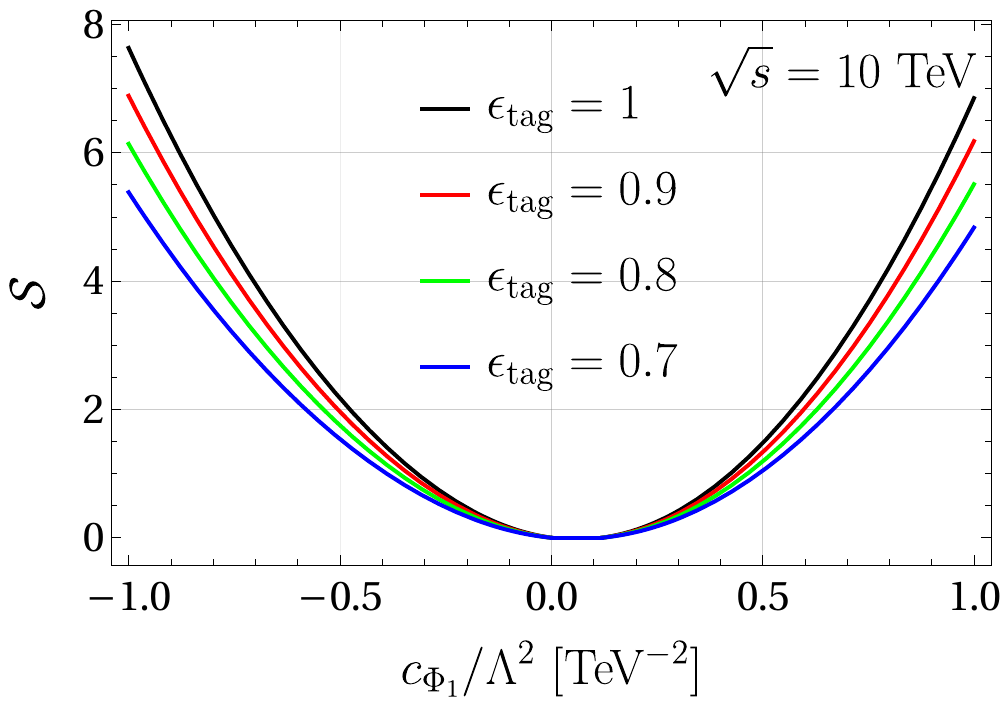}\\
\includegraphics[width=0.45\textwidth]{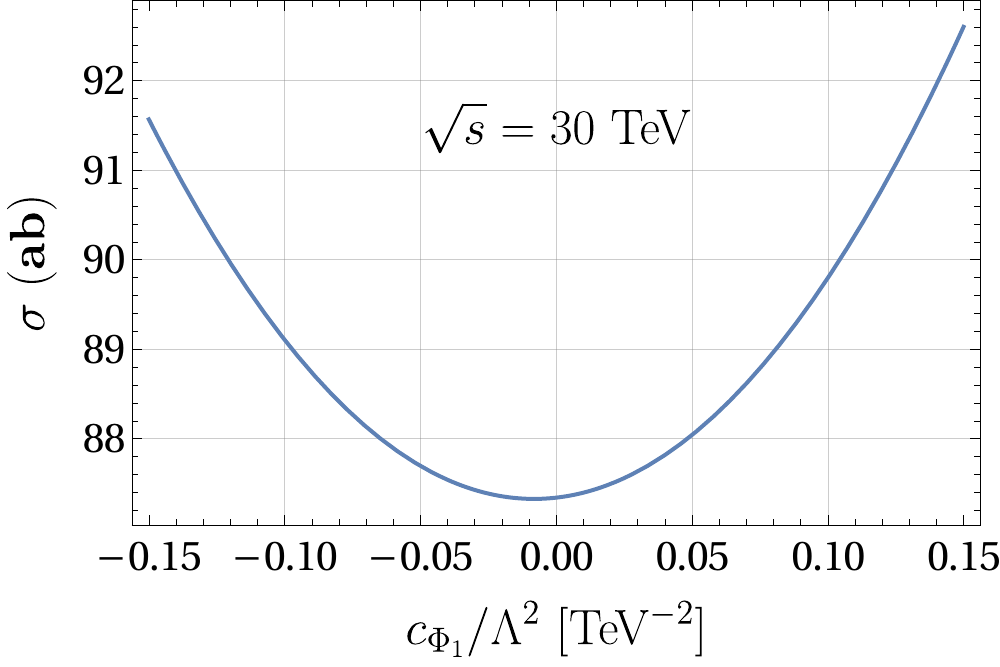}
\includegraphics[width=0.45\textwidth]{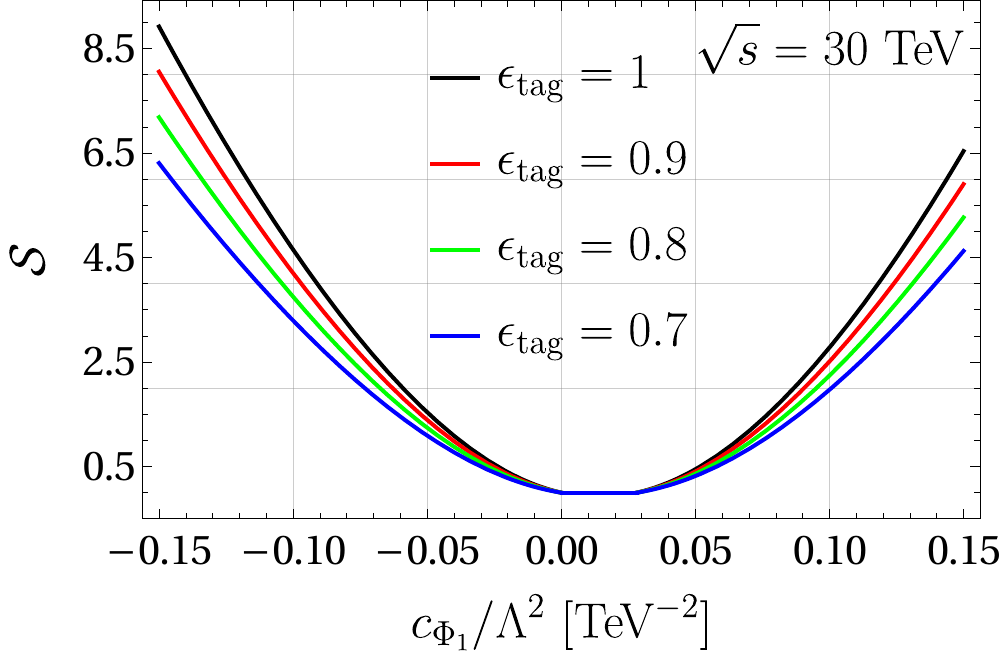}
\caption{The initial cross section of the $WWh$ signal without any cuts (left) and the significance $\mathcal{S}$ after combining the analyzed results (right), as a function of $c_{\Phi_1}/\Lambda^2$ for $\sqrt{s}=10$~TeV and $\mathcal{L}=10~\mathrm{ab}^{-1}$ (upper) as well as $\sqrt{s}=30$~TeV and $\mathcal{L}=90~\mathrm{ab}^{-1}$ (lower).}
\label{fig:wwh_significance_scan_cdp_10TeV}
\end{figure}

We further study the sensitivity of the parameter $c_{\Phi_1}$ at muon collider by fixing $c_6/\Lambda^2=0$ and performing a scan of $c_{\Phi_1}/\Lambda^2$ in the range $[-1,1]$ TeV$^{-2}$. The same cuts are applied as given above. We show the cross sections of the signal and the significance $\mathcal{S}$ as functions of $c_{\Phi_1}/\Lambda^2$ in Fig.~\ref{fig:wwh_significance_scan_cdp_10TeV}. We find that the cross section curve shifts to the left, while the significance curve shifts a little to the opposite direction due to the effect of cut efficiency. We can see that the $1\sigma$ and $2\sigma$ ranges of $c_{\Phi_1}/\Lambda^2$ are $[-0.318, 0.424]$ TeV$^{-2}$ and $[-0.477, 0.571]$ TeV$^{-2}$ for $\sqrt{s}=10$~TeV, respectively.
For $\sqrt{s}=30$~TeV, we also perform a scan of $c_{\Phi_1}/\Lambda^2$ in the range $[-0.15,0.15]$ TeV$^{-2}$,
and the $1\sigma$ and $2\sigma$ ranges of $c_{\Phi_1}/\Lambda^2$ are $[-0.0378, 0.0657]$ TeV$^{-2}$ and $[-0.0591, 0.0867]$ TeV$^{-2}$, respectively.

Finally, we comment on the other possible choice of signal. One might be tempted to choose the signal as $\ell^+\ell^-b\bar{b}$ plus large missing energy by instead decaying $W^\pm$ leptonically. However, in this case, the top quark in the background process $\mu^+ \mu^-\rightarrow \nu \bar{\nu} t \bar{t}$ cannot be fully reconstructed and thus the events are still much more than those of the signal without imposing the top veto cut. As a result, this process has no competitive reach of sensitivity.

\subsection{$WW\to ZZh$}
\label{sec:zzh}

\begin{figure}[th!]
\centering
\includegraphics[width=0.45\textwidth]{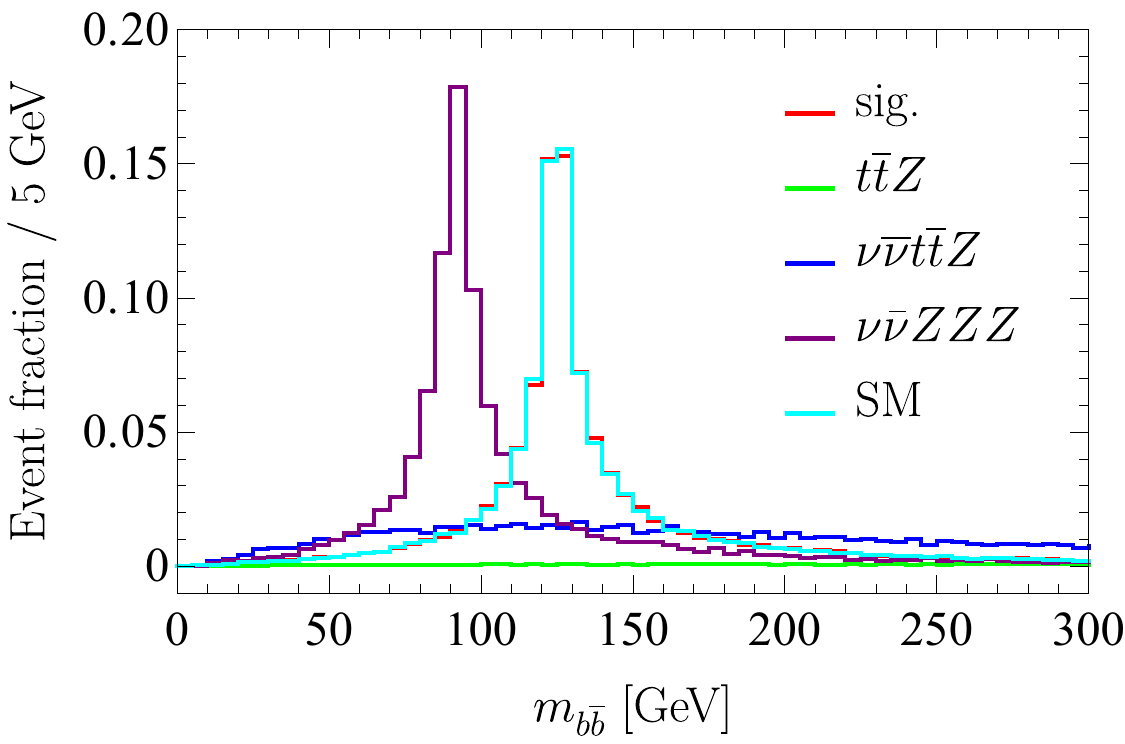}
\includegraphics[width=0.45\textwidth]{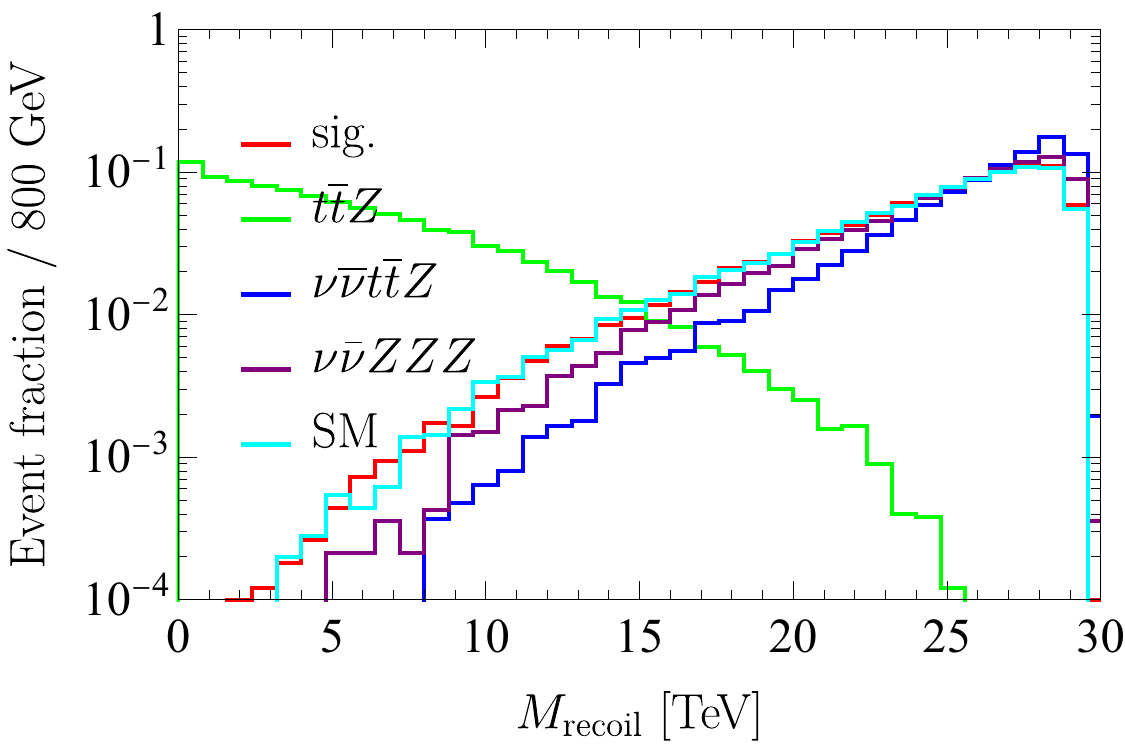}
\caption{The kinematic distributions of $ZZh$ signal and SM backgrounds for $\sqrt{s}=30$ TeV. Left: the invariant mass of the two $b$ jets $m_{b\bar{b}}$. Right: the recoil mass $M_{\rm recoil}$.}
\label{fig:zzh}
\end{figure}

Next we consider the following VBF topology for $ZZh$ production
\begin{eqnarray}
\mu^+ \mu^-\to \nu_\mu \bar{\nu}_\mu ZZ h~~~(WW~{\rm fusion})\;.
\end{eqnarray}
The final states are followed by the semileptonic decay channel of $Z$ bosons $ZZ\to \ell^+\ell^- jj$ as well as $h\to b\bar{b}$. The signal would be $\ell^+\ell^- jj b\bar{b}$ plus large missing energy, with the $b\bar{b}$~($\ell^+\ell^-$ and $jj$) pair being near the Higgs ($Z$) boson mass resonance.
The irreducible backgrounds thus consist of
\begin{eqnarray}
\mu^+\mu^-\to t\bar{t}Z\;,~~ \nu\bar{\nu}t\bar{t}Z\;,~~
\nu\bar{\nu} ZZZ\;.
\end{eqnarray}
In the first two backgrounds, the two top quarks leptonically decay and the $Z$ boson decays into two light jets. In the background process of $\nu\bar{\nu} ZZZ$, the three $Z$ bosons decay to $\ell^+\ell^-, j j$ and $b\bar{b}$, respectively. We consider $\sqrt{s}=30$~TeV~\footnote{For $\sqrt{s}=10$~TeV, we find the best achievable significance is less than one and the number of events after applying various cuts is too small.} and impose the basic acceptance cuts on the objects
\begin{eqnarray}\label{cut:zzh_pt}
p_T(\ell,b,j)>60~{\rm GeV}\;,~~~10^{\circ}<\theta_{\ell,b,j}<170^{\circ}\;.
\end{eqnarray}
For the signal events, we reconstruct the Higgs boson $h$ with the invariant mass of the two $b$ jets as shown in the left panel of Fig.~\ref{fig:zzh} and apply the invariant mass cut as follows
\begin{eqnarray}\label{cut:zzh_mbb}
m_{b\bar{b}}=m_h\pm 20~{\rm GeV}\;.
\end{eqnarray}
For the $Z$ bosons reconstructed from $\ell^+\ell^-$ and $jj$ pairs, the invariant mass cuts are not applied since these two cuts would further reduce the number of signal events as well as the significance.
We also take the recoil mass cut
\begin{equation}\label{cut:zzh_rec_mass}
M_{\text{recoil}}>15~\rm{TeV}
\end{equation}
to reduce the $t\bar{t}Z$ background as shown in the right panel of Fig.~\ref{fig:zzh}.

\begin{figure}[th!]
\centering
\includegraphics[width=0.45\textwidth]{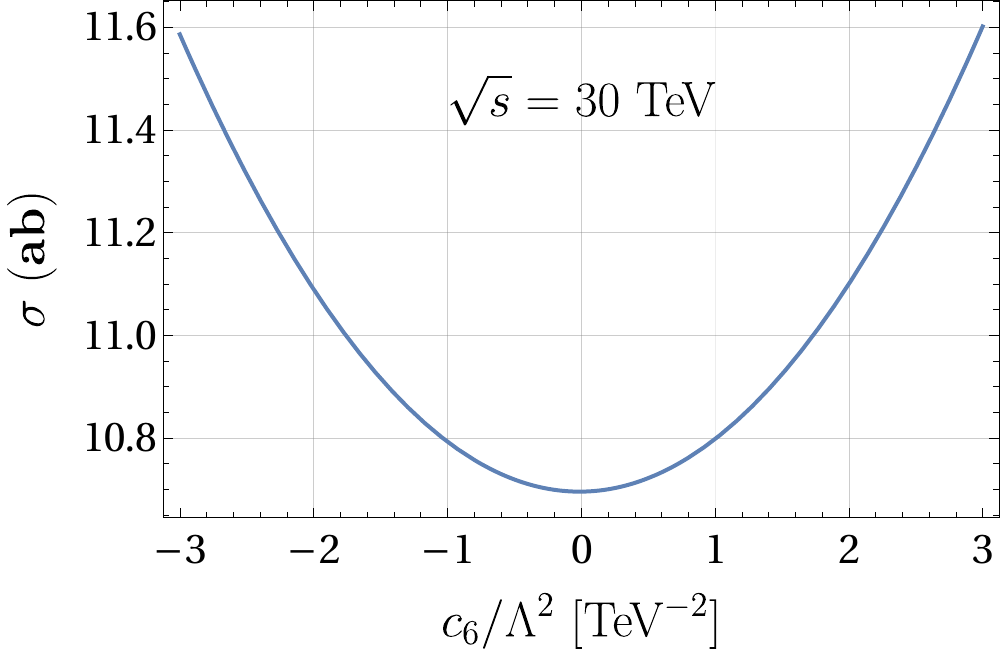}
\includegraphics[width=0.45\textwidth]{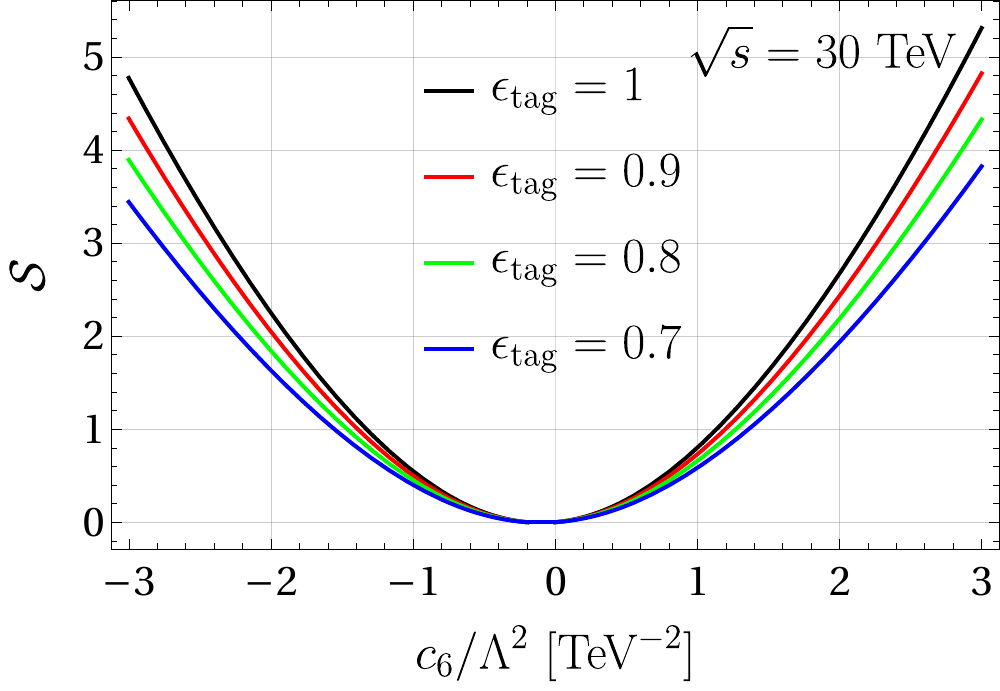}\\
\includegraphics[width=0.45\textwidth]{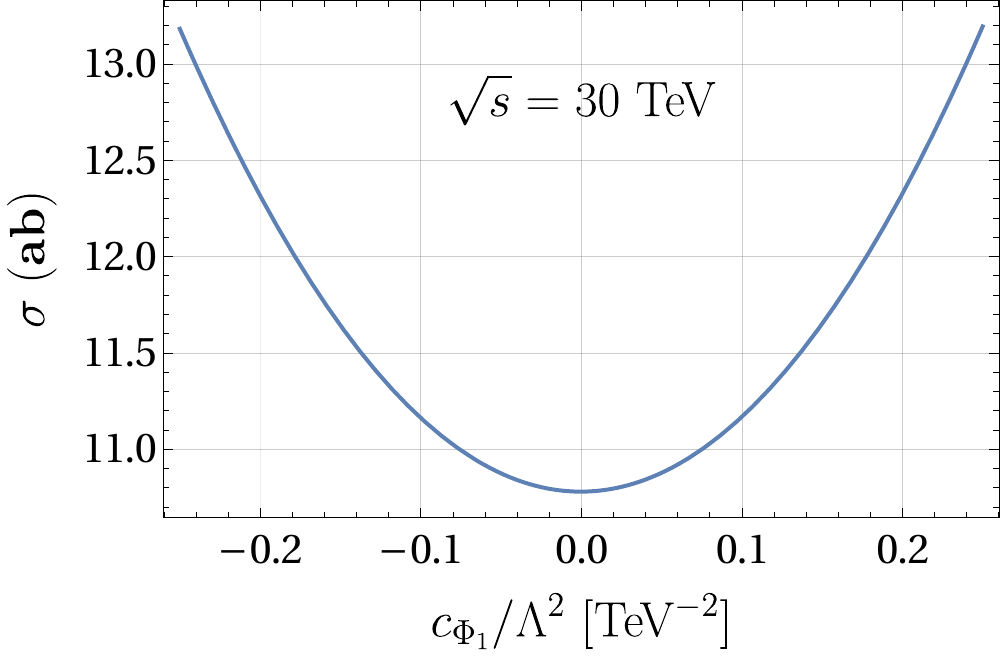}
\includegraphics[width=0.45\textwidth]{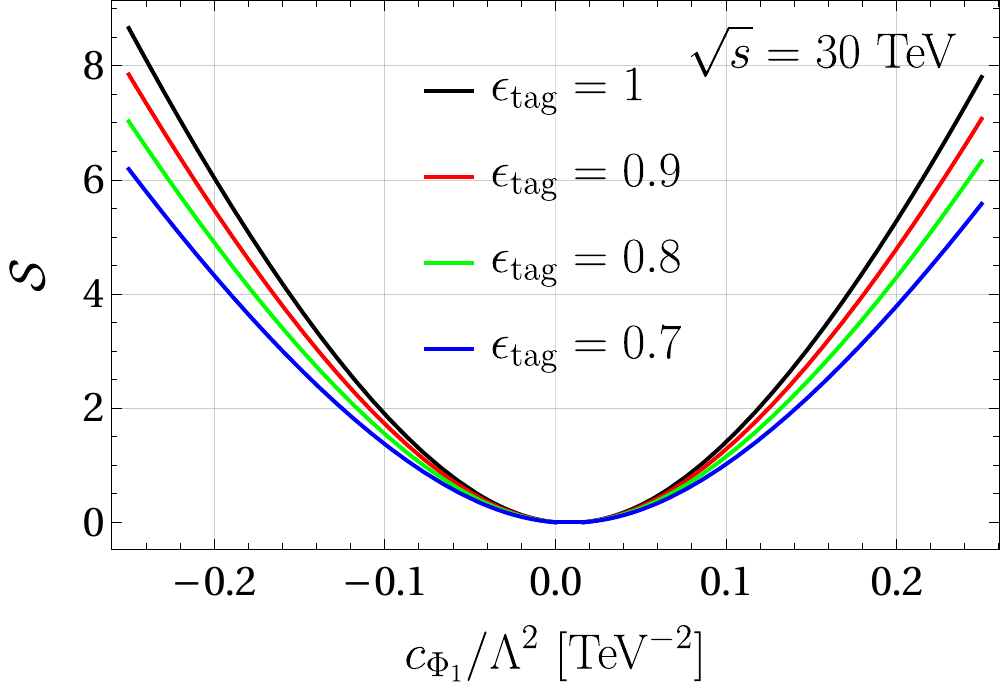}
\caption{The initial cross section of the $ZZh$ signal without any cuts (left) and the significance $\mathcal{S}$ after combining the analyzed results (right), as a function of $c_{6}/\Lambda^2$ (upper) and $c_{\Phi_1}/\Lambda^2$ (lower) for $\sqrt{s}=30$~TeV and $\mathcal{L}=90~\mathrm{ab}^{-1}$.}
\label{fig:zzh_significance}
\end{figure}

We then show the cut efficiencies and the number of events in Table~\ref{tab:zzh_cut_eff} by fixing $c_6/\Lambda^2=-2$ TeV$^{-2}$ and $c_{\Phi_1}/\Lambda^2=0$ for $\sqrt{s}=30$ TeV with the luminosity of 90 ab$^{-1}$.
For this setup one can obtain the significance as $\mathcal{S}=2.25$. We also fix $c_{\Phi_1}/\Lambda^2=0$ and perform a scan of the parameter $c_6/\Lambda^2$ in the range $[-3,3]$ TeV$^{-2}$ to evaluate the projected sensitivity to the effective coupling.
The cross section before the selection cuts of the $ZZh$ signal and the final significance $\mathcal{S}$ are displayed as a function of $c_6/\Lambda^2$ in the upper panels of Fig.~\ref{fig:zzh_significance}.
One can see that the $1\sigma$ and $2\sigma$ ranges of $c_6/\Lambda^2$ are $[-1.329, 1.136]$ TeV$^{-2}$ and $[-1.881, 1.691]$ TeV$^{-2}$, respectively.
We further fix $c_6/\Lambda^2=0$ and perform a scan of $c_{\Phi_1}/\Lambda^2$ in the range of $[-0.25,0.25]$ TeV$^{-2}$. The same cuts as given in above analysis are also applied. The lower panels of Fig.~\ref{fig:zzh_significance} show that the $1\sigma$ and $2\sigma$ ranges of $c_{\Phi_1}/\Lambda^2$ are $[-0.0688, 0.0852]$ TeV$^{-2}$ and $[-0.103, 0.119]$ TeV$^{-2}$, respectively.

\begin{table}[tb!]
\centering
\begin{tabular}{|c|c|c|c|c|c|}
\hline
$ZZh$ & \text{signal} & $t\bar{t}Z$ & $\nu\bar{\nu}t\bar{t}Z$ & $\nu\bar{\nu}ZZZ$ & $c_6=0$ (SM) \\ \hline
initial & 997.666 & 23.090 & 1671.074 & 458.515 & 961.092\\ \hline
$p_T$ and $\theta$ cut (Eq.~(\ref{cut:zzh_pt}))& 13.461 & 15.657 & 30.238 & 0.882 & 4.994 \\ \hline
$m_{b\bar{b}}$ cut (Eq.~(\ref{cut:zzh_mbb}))& 8.209 & 0.078 & 1.061 & 0.163 & 2.891 \\ \hline
$M_{\text{recoil}}$ cut (Eq.~(\ref{cut:zzh_rec_mass}))& 8.121 & 0.000 & 1.061 & 0.163 & 2.816 \\ \hline
Efficiency $\epsilon (\%)$ & 0.814 & 0.000 & 0.063 & 0.036 & 0.293 \\ \hline
\end{tabular}
\caption{The selection efficiencies for the $ZZh$ signal with $c_6/\Lambda^2=-2$ TeV$^{-2}$ and $c_{\Phi_1}/\Lambda^2=0$ and SM backgrounds. The number of events before and after selection cuts are evaluated at the muon collider with $\sqrt{s}=30$ TeV and luminosity $\mathcal{L}=90$ ab$^{-1}$. The number of signal events is too small after imposing necessary cuts at the muon collider with $\sqrt{s}=10$ TeV, thus the results are not listed here.}
\label{tab:zzh_cut_eff}
\end{table}

\subsection{$WW\to hhh$}
\label{sec:hhh}

For $hhh$ production we consider the following VBF topology
\begin{eqnarray}
\mu^+ \mu^-\to \nu_\mu \bar{\nu}_\mu hhh~~~(WW~{\rm fusion})\;.
\end{eqnarray}
The leading signal would be six $b$ jets as well as missing energy. The main background includes
\begin{eqnarray}
\mu^+ \mu^-\to \nu \bar{\nu} ZZZ\,,~~\nu \bar{\nu} ZZh\;,~~\nu \bar{\nu} Zhh\;,
\end{eqnarray}
with the $Z$ boson and the SM Higgs decaying into $b\bar{b}$ pairs. Compared with the signal, the cross sections of the background are suppressed by the branching fraction of $Z\to b\bar{b}$ which makes it possible to achieve a high significance with very loose cuts at muon collider. Note that there also exists the
continuum background from SM processes with multiple $b$ jets. Data-driven methods are usually used to estimate the continuum background contribution. The $b$ jets are required to satisfy a set of isolation criteria to reject the continuum background from
light jets misidentified as $b$ jets or from virtual photon $\gamma^\ast\to b\bar{b}$. We assume a subtraction of the continuum background
and neglect it in the following study. We consider $\sqrt{s}=10$~TeV and impose the basic acceptance cuts on the $b$ jets
\begin{eqnarray}
\label{eq:hhh_cut}\label{cut:hhh_pt}
p_T(b)>40~{\rm GeV}\;,~~~10^{\circ}<\theta_{b}<170^{\circ}\;.
\end{eqnarray}
In order to keep enough signal events, the invariant mass cut of the reconstructed Higgs boson is not applied. The recoil mass cut introduced in the above processes is also not applied in this case, since there is no significant difference between the signal and backgrounds.

\begin{figure}[th!]
\centering
\includegraphics[width=0.45\textwidth]{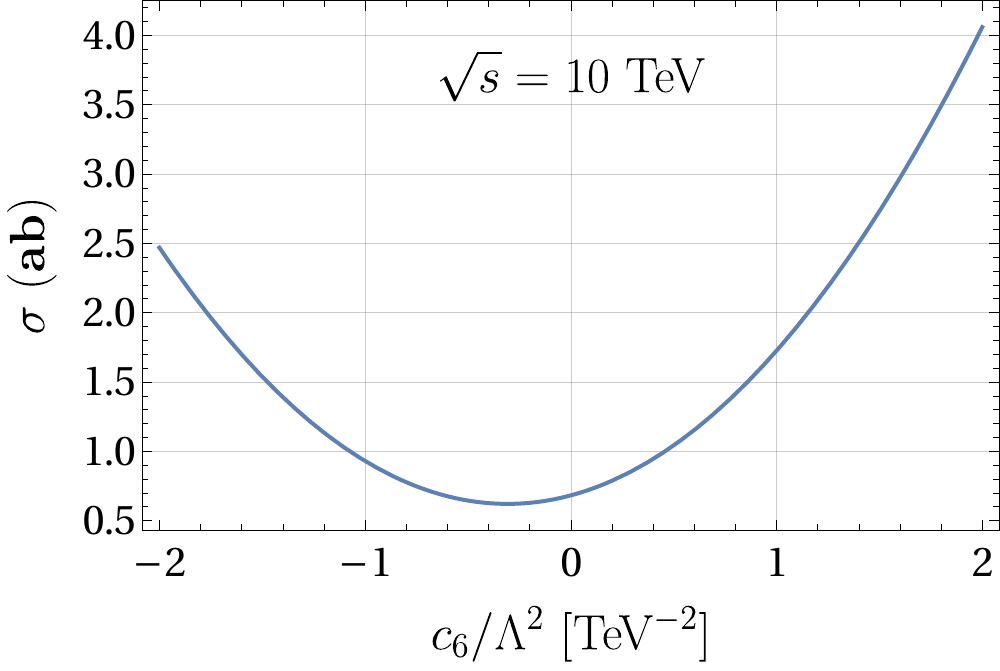}
\includegraphics[width=0.45\textwidth]{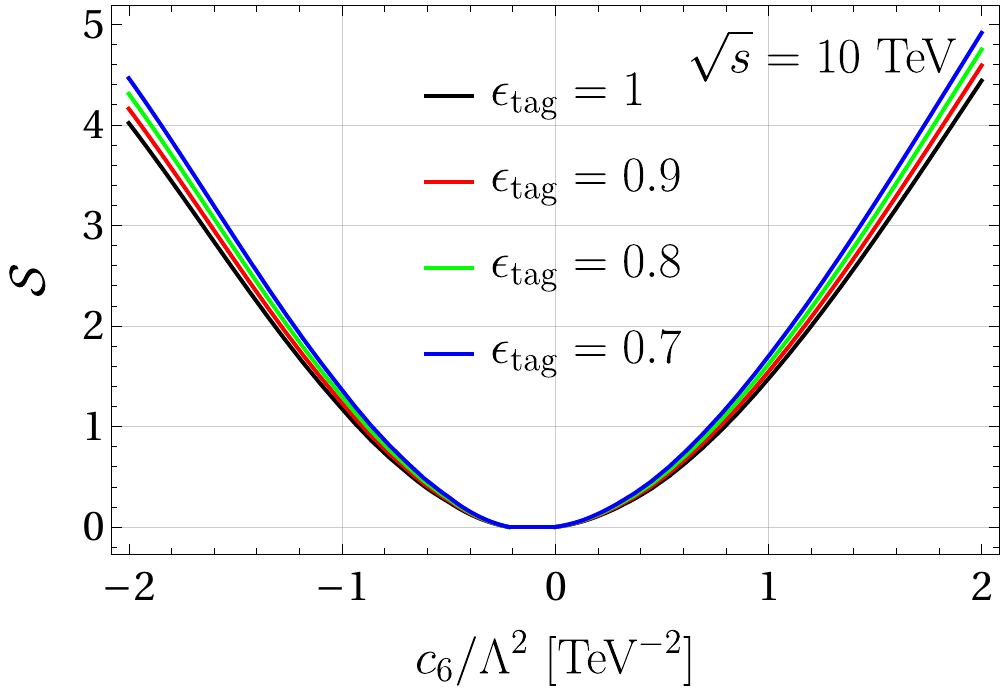}\\
\includegraphics[width=0.45\textwidth]{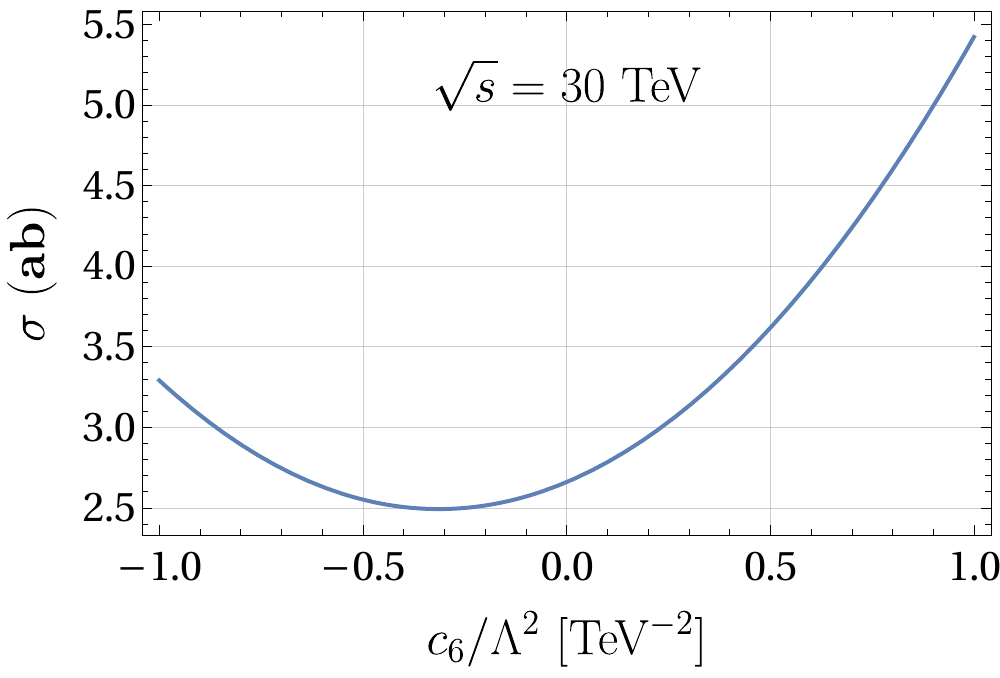}
\includegraphics[width=0.45\textwidth]{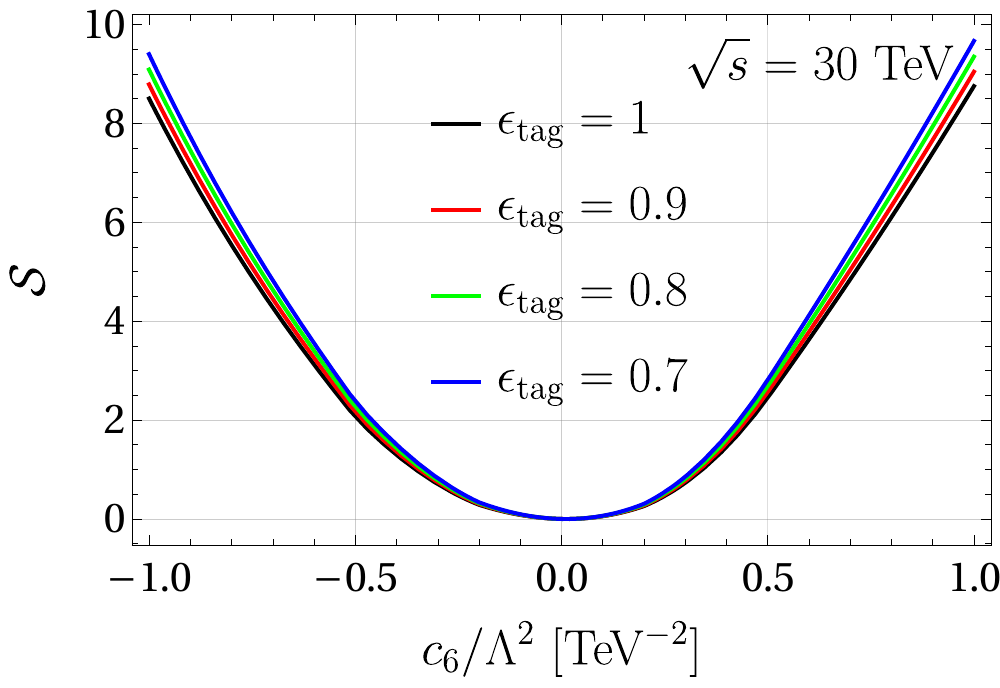}
\caption{The initial cross section of the $hhh$ signal without any cuts (left) and the significance $\mathcal{S}$ after combining the analyzed results (right), as a function of $c_6/\Lambda^2$ for $\sqrt{s}=10$~TeV and $\mathcal{L}=10~\mathrm{ab}^{-1}$ (upper) as well as $\sqrt{s}=30$~TeV and $\mathcal{L}=90~\mathrm{ab}^{-1}$ (lower).}
\label{fig:hhh_significance_cp}
\end{figure}

The cut efficiencies and the number of events are given outside the brackets in Table~\ref{tab:hhh_cut_eff_cp} for benchmark $c_6/\Lambda^2=-1$ TeV$^{-2}$ and $c_{\Phi_1}/\Lambda^2=0$ with $\sqrt{s}=10$ TeV and the luminosity of 10 ab$^{-1}$. For this setup one can obtain the significance as $\mathcal{S}=1.18$.
From the cross section and significance as a function of $c_6/\Lambda^2$ shown
in Fig.~\ref{fig:hhh_significance_cp} for $\sqrt{s}=10$ TeV, one can see that the $1\sigma$ and $2\sigma$ ranges of $c_6/\Lambda^2$ are $[-0.926, 0.796]$ TeV$^{-2}$ and $[-1.316, 1.201]$ TeV$^{-2}$, respectively.
For muon collider with $\sqrt{s}=30$~TeV, we instead require $p_T(b)>50~{\rm GeV}$. The results of cut efficiency are given inside the brackets in Table~\ref{tab:hhh_cut_eff_cp} for $\sqrt{s}=30$ TeV with the luminosity of 90 ab$^{-1}$ for benchmark $c_6/\Lambda^2=-0.5$ TeV$^{-2}$ and $c_{\Phi_1}/\Lambda^2=0$. This setup leads to the significance as $\mathcal{S}=2.06$. The corresponding results are also shown in Fig.~\ref{fig:hhh_significance_cp}, and the $1\sigma$ and $2\sigma$ ranges of $c_6/\Lambda^2$ are $[-0.354, 0.342]$ TeV$^{-2}$ and $[-0.493, 0.458]$ TeV$^{-2}$, respectively. It is interesting to note that the sensitivities of the cross sections to $c_6$ before cuts are larger for $c_6>0$ than for $c_6<0$, whereas the significance are symmetric for $c_6>0$ and $c_6<0$. The reason is that the cut efficiency has the opposite dependence on $c_6$ to the cross sections, which balances out the significance eventually. We should also note that in this process, higher tagging efficiencies of longitudinal modes would increase the number of background events and thus lead to a decreasing sensitivity that can be seen in the right panels of Fig.~\ref{fig:hhh_significance_cp}. This dependence is in opposite to other processes above.

\begin{figure}[th!]
\centering
\includegraphics[width=0.45\textwidth]{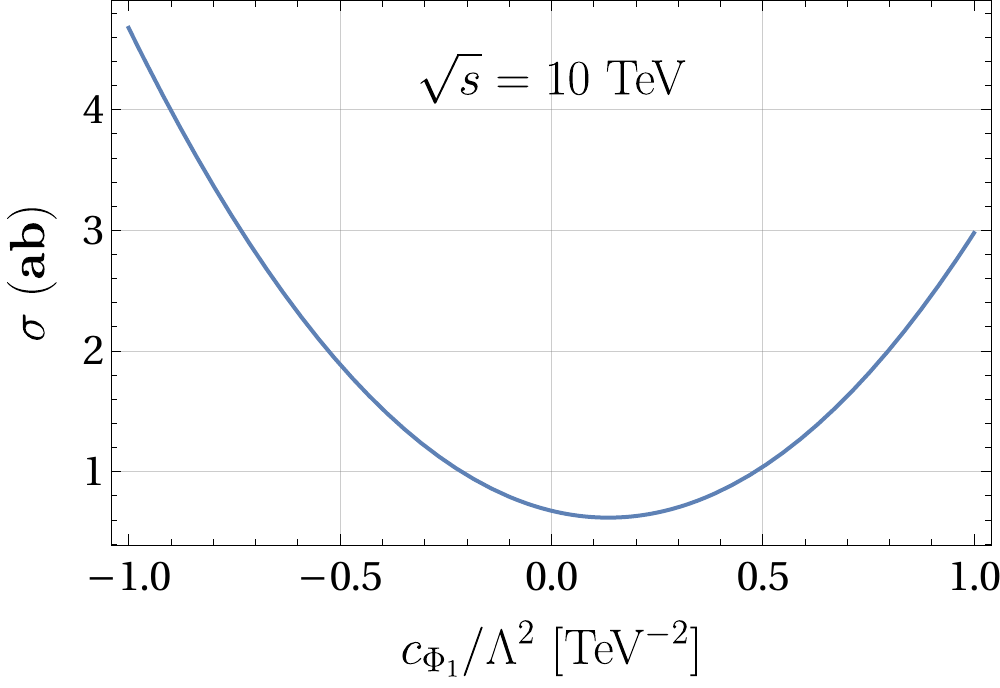}
\includegraphics[width=0.45\textwidth]{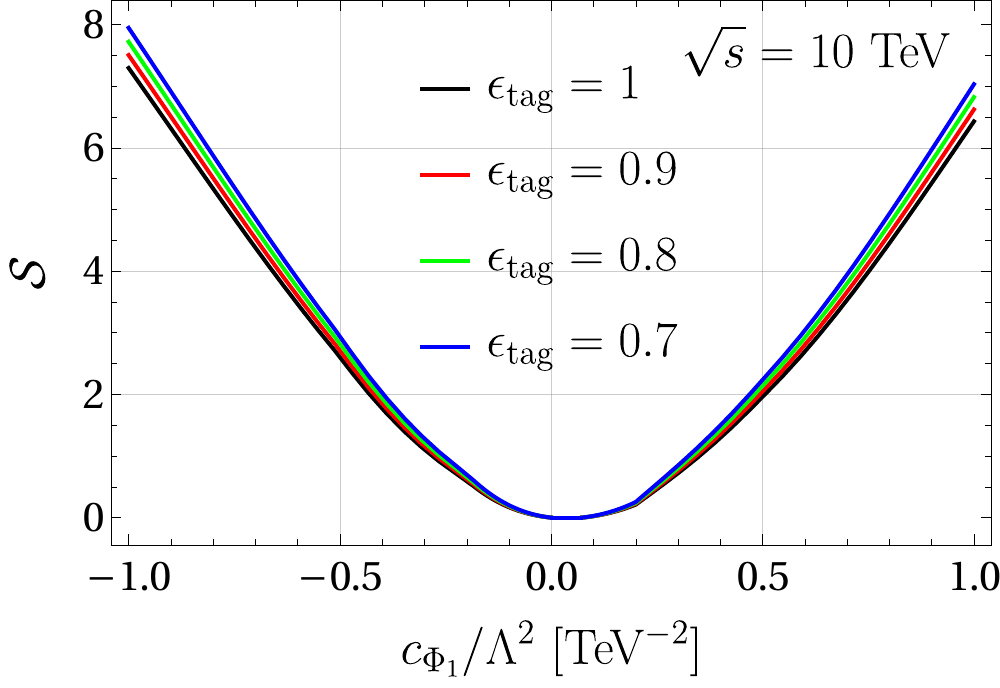}\\
\includegraphics[width=0.45\textwidth]{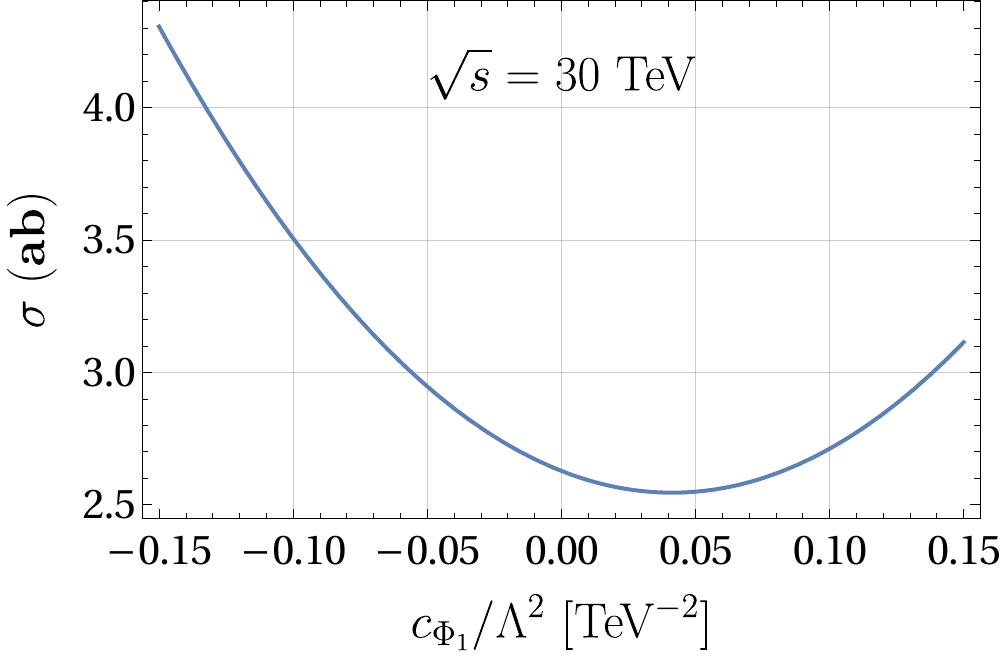}
\includegraphics[width=0.45\textwidth]{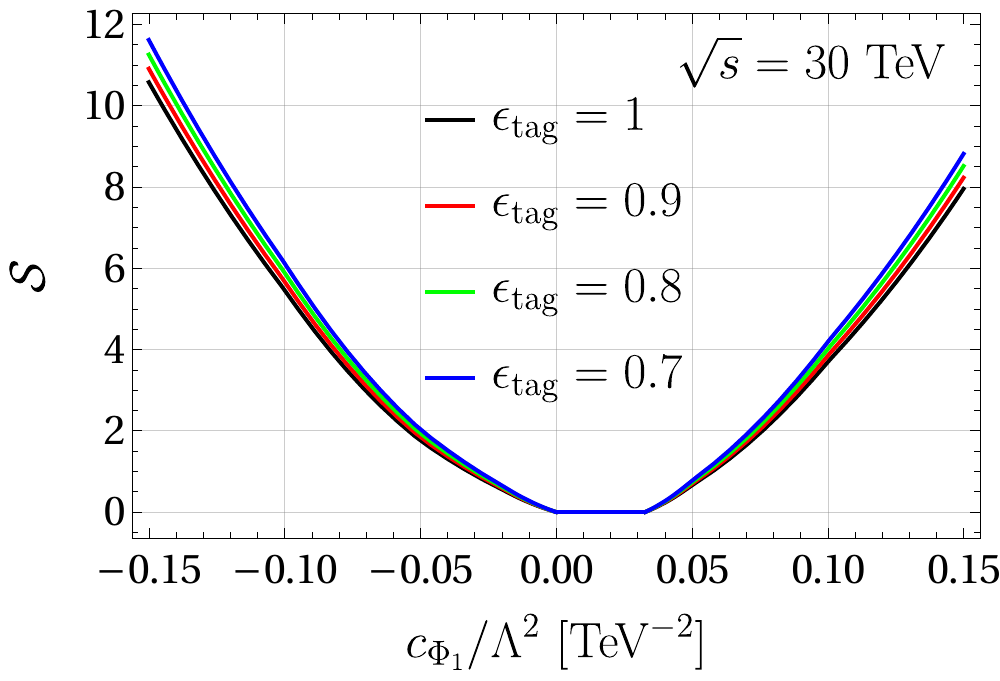}
\caption{The initial cross section of the $hhh$ signal without any cuts (left) and the significance $\mathcal{S}$ after combining the analyzed results (right), as a function of $c_{\Phi_1}/\Lambda^2$ for $\sqrt{s}=10$~TeV and $\mathcal{L}=10~\mathrm{ab}^{-1}$ (upper) as well as $\sqrt{s}=30$~TeV and $\mathcal{L}=90~\mathrm{ab}^{-1}$ (lower).}
\label{fig:hhh_significance_cdp}
\end{figure}

We further perform a scan of $c_{\Phi_1}/\Lambda^2$ in the range $[-1,1]$ TeV$^{-2}$ and apply the same cuts as given in Eq.~\eqref{eq:hhh_cut}. The cross section of the signal and the significance $\mathcal{S}$ as a function of $c_{\Phi_1}/\Lambda^2$ are displayed in Fig.~\ref{fig:hhh_significance_cdp} for both $\sqrt{s}=10$ TeV and 30 TeV. For $\sqrt{s}=10$ (30) TeV, the $1\sigma$ and $2\sigma$ ranges of $c_{\Phi_1}/\Lambda^2$ are $[-0.282, 0.351]~([-0.0324,0.0576])$ TeV$^{-2}$ and $[-0.430, 0.505]~([-0.0545,0.0760])$ TeV$^{-2}$, respectively.

\begin{table}[tb!]
\centering
\begin{tabular}{|c|c|c|c|c|c|}
\hline
$hhh$ & \text{signal} & $\nu\bar{\nu}ZZZ$ & $\nu\bar{\nu}ZZh$ & $\nu\bar{\nu}Zhh$ & $c_6=0$ (SM) \\ \hline
initial & 9.298 [229.556] & 3.498 [95.706] & 11.858 [334.314] & 16.023 [462.745] & 6.820 [239.323] \\ \hline
$p_T$ and $\theta$ cut & 2.027 [11.092] & 0.099 [0.419] & 0.466 [3.162] & 0.697 [5.248] & 0.326 [3.250] \\ \hline
Efficiency $\epsilon(\%)$ & 21.804 [4.832] & 2.830 [0.438] & 3.928 [0.946] & 4.352 [1.134] & 4.784 [1.358] \\ \hline
\end{tabular}
\caption{
The selection efficiencies for the $hhh$ signal and SM backgrounds. The number of events before and after selection cuts are evaluated with $c_6/\Lambda^2=-1$ TeV$^{-2}$ and $c_{\Phi_1}/\Lambda^2=0$ at muon collider with $\sqrt{s}=10$ TeV and the luminosity of $\mathcal{L}=10$ ab$^{-1}$ (outside brackets), and with $c_6/\Lambda^2=-0.5$ TeV$^{-2}$ and $c_{\Phi_1}/\Lambda^2=0$ at muon collider with $\sqrt{s}=30$ TeV and the luminosity of $\mathcal{L}=90$ ab$^{-1}$ (inside brackets).}
\label{tab:hhh_cut_eff_cp}
\end{table}

\subsection{Summary}
\label{sec:combine}

We collected all projected sensitivities to the coupling $c_6/\Lambda^2$ and $c_{\Phi_1}/\Lambda^2$ from the above three processes in Table~\ref{tab:summary_table_coupling}. As expected, the muon collider with higher energy and larger luminosity exhibits a better sensitivity for the couplings. For all processes, the sensitivity for $c_{\Phi_1}/\Lambda^2$ is better than that for $c_{6}/\Lambda^2$. The expected sensitivity to the coefficients $c_6/\Lambda^2$ and $c_{\Phi_1}/\Lambda^2$ can reach the level of 0.01 TeV$^{-2}$.
We also summarize the energy scale which these processes can reach in Table~\ref{tab:summary_table_scale}.

\begin{table}[th!]
\centering
\begin{tabular}{|c|c|c|c|c|}
\hline
channels & $\sqrt{s}$ & coupling $[\text{TeV}^{-2}]$ & $1\sigma$ & $2\sigma$ \\ \hline
\multirow{4}{*}{$WWh$} & \multirow{2}{*}{10 TeV} &
$c_6/\Lambda^2$ & $[-0.856,0.940]$ & $[-1.245,1.327]$\\ \cline{3-5}
& & $c_{\Phi_1}/\Lambda^2$ & $[-0.318, 0.424]$ & $[-0.477,0.571]$\\ \cline{2-5}
& \multirow{2}{*}{30 TeV} &
$c_6/\Lambda^2$ & $[-0.447,0.389]$ & $[-0.627,0.569]$\\ \cline{3-5}
& & $c_{\Phi_1}/\Lambda^2$ & $[-0.0378, 0.0657]$ & $[-0.0591, 0.0867]$\\ \hline
\multirow{2}{*}{$ZZh$} & \multirow{2}{*}{30 TeV} &
$c_6/\Lambda^2$ & $[-1.329, 1.136]$ & $[-1.881, 1.691]$\\ \cline{3-5}
& & $c_{\Phi_1}/\Lambda^2$ & $[-0.0688, 0.0852]$ & $[-0.103, 0.119]$\\ \hline
\multirow{4}{*}{$hhh$} & \multirow{2}{*}{10 TeV} &
$c_6/\Lambda^2$ & $[-0.926, 0.796]$ & $[-1.316, 1.201]$\\ \cline{3-5}
& & $c_{\Phi_1}/\Lambda^2$ & $[-0.282, 0.351]$ & $[-0.430, 0.505]$\\ \cline{2-5}
& \multirow{2}{*}{30 TeV} &
$c_6/\Lambda^2$ & $[-0.354, 0.342]$ & $[-0.493, 0.458]$\\ \cline{3-5}
& & $c_{\Phi_1}/\Lambda^2$ & $[-0.0324, 0.0576]$ & $[-0.0545, 0.0760]$\\ \hline
\end{tabular}
\caption{The summary table of the expected sensitivities to the couplings at $1\sigma$ and $2\sigma$ for the three processes $WWh$, $ZZh$, and $hhh$ for $\sqrt{s}=10$~TeV and $\sqrt{s}=30$~TeV. The tagging efficiency of longitudinal polarizations is assumed to be $100\%$.
}
\label{tab:summary_table_coupling}
\end{table}

\begin{table}[th!]
\centering
\begin{tabular}{|c|c|c|c|c|}
\hline
channels & $\sqrt{s}$ & energy scales $[\text{TeV}]$ & $1\sigma$ & $2\sigma$ \\ \hline
\multirow{4}{*}{$WWh$} & \multirow{2}{*}{10 TeV} &
$\Lambda/\sqrt{c_6}$ & $1.032$ & $0.868$\\ \cline{3-5}
& & $\Lambda/\sqrt{c_{\Phi_1}}$ & $1.535$ & $1.323$\\ \cline{2-5}
& \multirow{2}{*}{30 TeV} &
$\Lambda/\sqrt{c_6}$ & $1.495$ & $1.263$\\ \cline{3-5}
& & $\Lambda/\sqrt{c_{\Phi_1}}$ & $3.899$ & $3.396$\\ \hline
\multirow{2}{*}{$ZZh$} & \multirow{2}{*}{30 TeV} &
$\Lambda/\sqrt{c_6}$ & $0.867$ & $0.729$\\ \cline{3-5}
& & $\Lambda/\sqrt{c_{\Phi_1}}$ & $3.427$ & $2.894$\\ \hline
\multirow{4}{*}{$hhh$} & \multirow{2}{*}{10 TeV} &
$\Lambda/\sqrt{c_6}$ & $1.039$ & $0.872$\\ \cline{3-5}
& & $\Lambda/\sqrt{c_{\Phi_1}}$ & $1.689$ & $1.407$\\ \cline{2-5}
& \multirow{2}{*}{30 TeV} &
$\Lambda/\sqrt{c_6}$ & $1.682$ & $1.425$\\ \cline{3-5}
& & $\Lambda/\sqrt{c_{\Phi_1}}$ & $4.166$ & $3.627$\\ \hline
\end{tabular}
\caption{The summary table of the reachable energy scales at $1\sigma$ and $2\sigma$ for the three processes $WWh$, $ZZh$, and $hhh$ for $\sqrt{s}=10$~TeV and $\sqrt{s}=30$~TeV. The tagging efficiency of longitudinal polarizations is assumed to be $100\%$.}
\label{tab:summary_table_scale}
\end{table}

\begin{figure}[th!]
\centering
\includegraphics[width=0.48\textwidth]{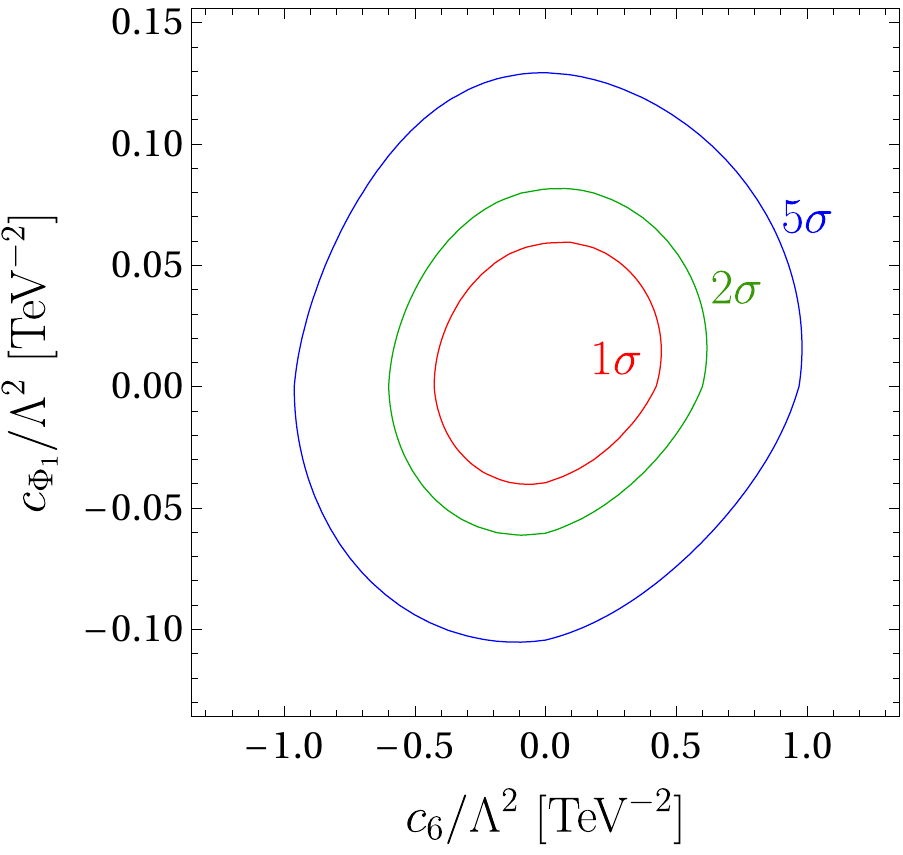}
\includegraphics[width=0.48\textwidth]{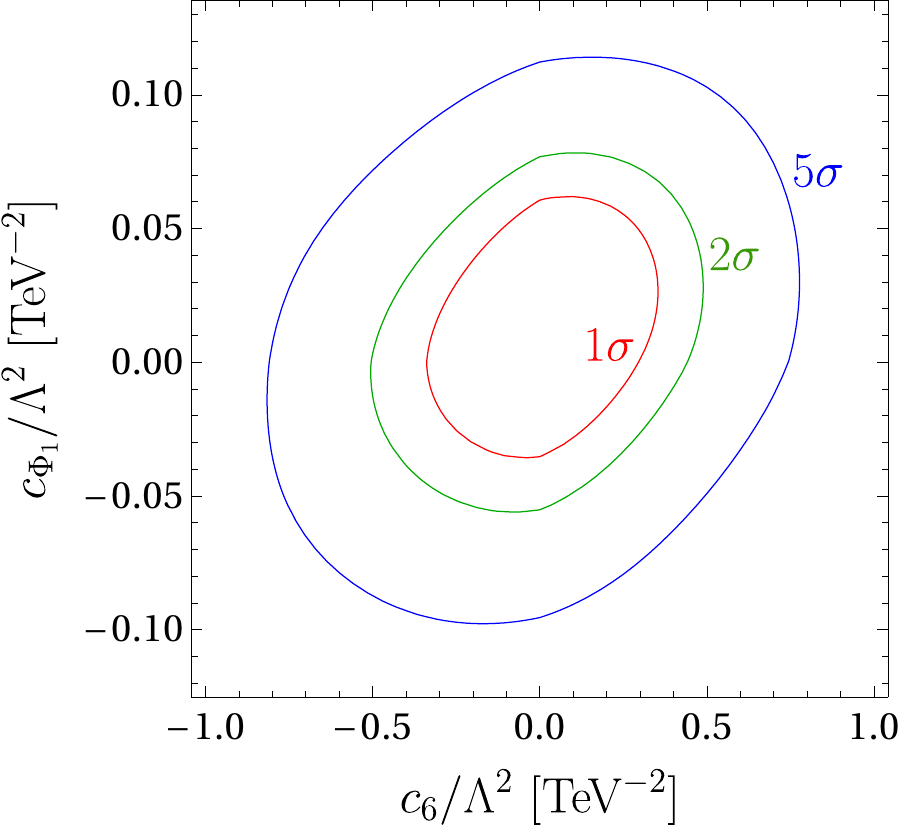}
\caption{The contour plots of the significance $\mathcal{S}$ in the plane of $c_{6}/\Lambda^2$ versus $c_{\Phi_1}/\Lambda^2$ from $WW\rightarrow WWh$ process (left) and $WW\rightarrow hhh$ process (right) for $\sqrt{s}=30$~TeV and $\mathcal{L}=90~\mathrm{ab}^{-1}$. The $ZZh$ process is not quite promising so we do not show the contour plot for it here.}
\label{fig:wwh_significance_contour_30TeV}
\end{figure}

To display the dependence of sensitivity on both $c_6/\Lambda^2$ and $c_{\Phi_1}/\Lambda^2$, we perform a scan in the plane of $(c_6/\Lambda^2, c_{\Phi_1}/\Lambda^2)$. We work at the muon collider with $\sqrt{s}=30$~TeV and $\mathcal{L}=90~\mathrm{ab}^{-1}$, and show the contour plots from $WW\rightarrow WWh$ process (left) and $WW\rightarrow hhh$ process (right) in Fig.~\ref{fig:wwh_significance_contour_30TeV}. The $1\sigma$, $2\sigma$ and $5\sigma$ significance contours are indicated with red, green, and blue curves, respectively.

\section{Conclusions and Discussions}
\label{sec:Con}

In this work we studied the measurement of Higgs self-couplings through $2\rightarrow 3$ VBS in future muon colliders with detailed background analysis, and obtained the constraints on Wilson coefficients $c_6/\Lambda^2$ and $c_{\Phi_1}/\Lambda^2$ as defined in Eq.~(\ref{eq:EFT_operator}).
We demonstrated that the $2\rightarrow 3$ VBS are  plausible channels for measuring the Higgs self-couplings at high-energy colliders. To ensure the sensitivities of VBS cross sections to the BSM physics at high energies, the involved gauge bosons are required chosen to be longitudinally polarized and high $p_T$ cuts are applied on the final states. We only implemented the $WW$ scatterings, i.e., $\mu^+\mu^-\rightarrow \nu \bar{\nu} VVh$ with $VVh$ being $WWh$, $ZZh$, and $hhh$, as their contributions to the cross sections dominate over the scattering of the neutral bosons.
In every channel, by assuming the tagging efficiency for the polarizations of final vector bosons to be $100\%$, we obtained the constraints on $c_6/\Lambda^2$, as well as $c_{\Phi_1}/\Lambda^2$, for $\sqrt{s}=10$ TeV and 30 TeV.
These results are summarized in Table~\ref{tab:summary_table_coupling} (see also Table~\ref{tab:summary_table_scale} for the reachable energy scales). In addition, we made contour plots for the constraints in the plane of $c_6/\Lambda^2$ versus $c_{\Phi_1}/\Lambda^2$ to display the correlation.

Our results confirmed the rough estimation of Refs.~\cite{Henning:2018kys,Chen:2021rid} that the $2\rightarrow 3$ VBS provides supplemental channels of measuring Higgs self-couplings besides the di-Higgs production at colliders. The tightest constraints come from the $hhh$ channel, whereas the $WWh$ channel is also comparable. The constraints from $ZZh$ are the weakest as the cross section is very small. If only considering $hhh$ channel, the constraints on $c_6/\Lambda^2$ at $1\sigma$ level are $[-0.926,0.796]~\text{TeV}^{-2}$ for $\sqrt{s}=10$ TeV with integrated luminosity $\mathcal{L}=10~\text{ab}^{-1}$, as well as $[-0.354,0.342] \ \text{TeV}^{-2}$ for $\sqrt{s}=30$ TeV with integrated luminosity $\mathcal{L}=90  \ \text{ab}^{-1}$. The constraints on $c_{\Phi_1}/\Lambda^2$ become $[-0.282,0.351]\ \text{TeV}^{-2}$ and $[-0.0324,0.0576]\  \text{TeV}^{-2}$ for $\sqrt{s}=10$ with $\mathcal{L}=10 \ \text{ab}^{-1}$ and $30$ TeV with $\mathcal{L}=90 \  \text{ab}^{-1}$, respectively.

The results we obtained from $WWh$ and $ZZh$ channels in Table~\ref{tab:summary_table_coupling} depend on the assumption of the polarization tagging efficiency being $100\%$. Since the cross sections with transverse $W$ or $Z$ bosons are much larger than those with longitudinal vector bosons, the correctly chosen polarization is necessary for our method. Fortunately, the efficiencies of tagging polarizations of $W/Z$ have been measured in other processes and can reach up to $80-99\%$.
Thus, we also made the plots to show the change of significance with the scan of tagging efficiency. The results shows that the significance remains consistently high when the tagging efficiency varies within the range of $[0.7,1]$.

Our analysis also leaves a lot of room for improvement. For example, it should be possible to construct more delicate observables to make use of the differential distribution of the cross sections. Besides, the background analysis is expected to be quite different in hadron colliders, thus warranting another separate study. In addition, although the cross sections of other channels (such as $W$ and $Z$) and other final states for the channels studied here have smaller cross sections, they are still worth  further study.
Finally, the focus in our work was the measurement of Higgs self-couplings and thus we exclusively studied $V_LV_L$ channels. However, the scattering of other polarizations can be used to measure and constrain the dim-6 operators in SMEFT other than $\mathcal{O}_6$ and $\mathcal{O}_{\Phi_1}$. We leave these topics for the research in future.

\acknowledgments
T.L. is supported by the National Natural Science Foundation of China (Grants No. 11975129, No. 12035008) and ``the Fundamental Research Funds for the Central Universities'', Nankai University (Grants No. 63196013).
C.Y.Y. is supported in part by the Grants No.~NSFC-11975130, No.~NSFC-12035008, No.~NSFC-12047533, by the National Key Research and Development Program of China under Grant No.~2017YFA0402200 and the China Post-doctoral Science Foundation under Grant No.~2018M641621.
J.M.C. is supported by Fundamental Research Funds for the Central Universities of China
No. 11620330. C.T.L. was supported by the KIAS Individual Grant No. PG075302 at Korea Institute for Advanced Study.
Y.W. thanks the U.S. Department of Energy for the financial support, under grant No. DE-SC 0016013.

\newpage

\bibliography{reference}

\end{document}